\documentclass{article}                                                                           

\usepackage[margin=1.3in]{geometry}                                                     

\usepackage{setspace}                                                                                          

\usepackage{indentfirst}                                                                          
\renewcommand\thesection{\Roman{section}}                                         
\renewcommand\thesubsection{\thesection.\Alph{subsection}}                   
\renewcommand\thesubsubsection{\thesubsection.\arabic{subsubsection}} 

\usepackage[explicit]{titlesec}                                                                  
\titleformat{\section}{\bfseries}{\thesection.}{1em}{\MakeUppercase{#1}}  
\titleformat{\subsection}{\bfseries}{\thesubsection.}{1em}{#1}                   
\titleformat{\subsubsection}{\itshape}{\thesubsubsection.}{1em}{#1}          

\usepackage{lastpage}                                                                             
\usepackage[figure,table]{totalcount}                                                        

\usepackage[]{footmisc}                                                                          

\usepackage{caption}                                                                              
\usepackage[labelformat=simple]{subcaption}                                           
\usepackage{standalone}
\usepackage{xifthen}     

\usepackage[english]{babel}	
\usepackage[utf8]{inputenc} 
\usepackage[kerning,spacing,babel]{microtype} 
\usepackage{eepic}      
\usepackage{epic}       
\usepackage[T1]{fontenc} 
\usepackage{lmodern}

\usepackage{authblk}		

\usepackage{upgreek}		
\usepackage[makeroom]{cancel}			

\usepackage{isotope}		
\usepackage[version=3]{mhchem}	

\usepackage{xspace}			
\usepackage{icomma}			
\usepackage{chngpage}		
\usepackage[usenames,dvipsnames,svgnames,table]{xcolor}			
\usepackage{enumitem}       
\usepackage{footnote}       

\usepackage[singlelinecheck=false, labelsep=quad]{caption}	
\usepackage{subcaption}		
\usepackage{placeins}   	
\usepackage{float}      	

\usepackage{supertabular}	
\usepackage{booktabs}		
\usepackage{array}			
\usepackage{multirow}		

\usepackage{graphicx}		
\usepackage{wrapfig}		
\usepackage{pgfplots}		
\usepackage{siunitx}

\usepackage{algorithmicx} 
\usepackage{listings} 	
\usepackage{verbatim}   

\usepackage{url}        
\usepackage{import}			
\usepackage{afterpage}  
\usepackage{lscape}     
\usepackage{rotating}   
\usepackage{chngcntr}   
\usepackage{appendix}   
\usepackage{titlesec}   

\usepackage{epsfig}     
\usepackage{epsf}       

\usepackage{hyperref}		

\usepackage{geometry}		
\usepackage{algorithm}  

\usepackage{amsmath}    
\usepackage{amsfonts}   
\usepackage{amssymb}    %
\usepackage{amsthm}     

\usepackage{nameref}		
\usepackage[capitalize,nameinlink]{cleveref}   

\makeatletter
\g@addto@macro\@floatboxreset\centering
\newcommand*{\currentname}{\@currentlabelname}
\makeatother

\newlength{\figheight}
\hypersetup{
  linktocpage=true,       
  bookmarks=true,         
  unicode=true,           
  pdftoolbar=true,        
  pdfmenubar=true,        
  pdffitwindow=false,     
  pdfstartview={FitH},    
  pdfnewwindow=true,      
  colorlinks=false,       
  linkcolor=red,          
  citecolor=green,        
  filecolor=magenta,      
  urlcolor=cyan,          
  pdfborder={0 0 0},      
}

\pgfplotsset{compat=1.14}

\lstdefinestyle{cpp}{
  belowcaptionskip=1\baselineskip,
  breaklines=true,
  frame=TB,
  xleftmargin=\parindent,
  language=C,
  showstringspaces=false,
  basicstyle=\footnotesize\ttfamily,
  keywordstyle=\bfseries\color{green!40!black},
  commentstyle=\itshape\color{purple!40!black},
  identifierstyle=\color{blue},
  stringstyle=\color{orange},
}

\lstdefinestyle{python}{
  belowcaptionskip=1\baselineskip,
  breaklines=true,
  frame=,
  xleftmargin=\parindent,
  language=Python,
  showstringspaces=false,
  basicstyle=\footnotesize\ttfamily,
  keywordstyle=\bfseries\color{blue},
  commentstyle=\itshape\color{gray},
  identifierstyle=\color{black},
  stringstyle=\color{green},
}

\lstdefinelanguage{Moose}{
  morekeywords={one,two,three,four,five,six,seven,eight,
  nine,ten,eleven,twelve,o,clock,rock,around,the,tonight},
  sensitive=false,
  morecomment=[l]{//},
  morecomment=[s]{/*}{*/},
  morestring=[b]",
}

\lstdefinestyle{moose}{
  belowcaptionskip=1\baselineskip,
  breaklines=true,
  frame=tb,
  xleftmargin=\parindent,
  language=Moose,
  showstringspaces=false,
  basicstyle=\footnotesize\ttfamily,
  keywordstyle=\bfseries\color{blue},
  commentstyle=\itshape\color{gray},
  identifierstyle=\color{black},
  stringstyle=\color{green},
}

\newcommand{\rattlesnake}{Rattlesnake\xspace}
\newcommand{\saaft}{SAAF\(\tau\)\xspace}

\newcommand{\parenthesis}[1]{{\left(#1 \right)}}
\newcommand{\bracket}[1]{{\left[ #1 \right]}}

\newcommand{\img}{\ensuremath{\hat{\imath}}\xspace}

\newcommand{\del}{\vec{\nabla}}

\newcommand{\inv}[1]{{{#1}^{-1}}}

\newcommand{\iter}[1]{^{#1}}
\newcommand{\error}[1]{\ensuremath{\delta #1}}

\newcommand{\e}[1]{\mathrm{e}^{#1}}
\newcommand{\tento}[1]{\ensuremath{10^{#1}}\xspace}

\newcommand{\half}[1][1]{\frac{#1}{2}}

\newcommand{\mat}[1]{\ensuremath{\mathbf{#1}}}
\newcommand{\tensor}[1]{\ensuremath{\underline{\underline{#1}}}}

\newcommand{\dd}[2][]{\frac{\partial #1}{\partial #2}}
\newcommand{\ddx}[1][]{\dd[#1]{x}}

\newcommand{\ddd}[2][{}]{\frac{\partial^2 #1}{\partial {#2}^2}}
\newcommand{\ddxx}[1][]{\ddd[#1]{x}}

\newcommand{\dx}[1][x]{\,d#1}

\newcommand{\dmu}{\dx[\mu]}
\newcommand{\domg}{\dx[\direction]}

\newcommand{\intsp}{\int_{4\pi}}

\newcommand{\intpolar}{\int_{-1}^{1}}

\newcommand*{\domain}{\ensuremath{\mathcal{D}}\xspace}
\newcommand*{\boundary}{\ensuremath{{\partial\domain}}\xspace}

\newcommand{\testfct}{\ensuremath{\phi^{*}}\xspace}
\newcommand{\atestfct}{\ensuremath{\psi^{*}}\xspace}
\newcommand{\normal}{\ensuremath{\vec{n}}\xspace}

\newcommand{\sn}[1][N]{\ensuremath{S_#1}\xspace}

\newcommand{\keff}{\ensuremath{k_{\text{eff}}}\xspace}

\newcommand{\addindex}[1]{\ifthenelse{\isempty{#1}}{}{,{#1}}}
\newcommand{\spharmonic}{\ensuremath{\mathrm{Y}_{l}^{p}\xspace}}
\newcommand{\mflux}[1][]{\ensuremath{\phi_{l\addindex{#1}}^{p}\xspace}}
\newcommand{\sigl}[2][{}]{\ensuremath{\sigma_{#2\addindex{#1}}}\xspace}

\newcommand{\addgroup}[1]{\ifthenelse{\isempty{#1}}{}{_{#1}}}
\newcommand{\direction}{\ensuremath{\vec{\Omega}}\xspace}

\newcommand{\current}[1][]{\ensuremath{\vec{J}\addgroup{#1}}\xspace}

\newcommand{\drift}[1][]{\ensuremath{\hat{D}\addgroup{#1}}\xspace}
\newcommand{\DC}[1][]{\ensuremath{\mathrm{D}\addgroup{#1}}\xspace}
\newcommand{\DCNL}[1][]{\ensuremath{\tensor{\mathrm{D}}\addgroup{#1}}\xspace}
\newcommand{\xslabel}[2][]{\ifthenelse{\isempty{#1}}{\mathrm{#2}}{\mathrm{#2},#1}}
\newcommand{\sigt}[1][]{\ensuremath{\sigma_{\xslabel[#1]{t}}}\xspace}
\newcommand{\sigs}[1][]{\ensuremath{\sigma_{\xslabel[#1]{s}}}\xspace}
\newcommand{\sigf}[1][]{\ensuremath{\sigma_{\xslabel[#1]{f}}}\xspace}

\newcommand{\siga}[1][]{\ensuremath{\sigma_{\xslabel[#1]{a}}}\xspace}

\newcommand{\weight}[1][]{\ensuremath{\mathrm{w}\addgroup{#1}}\xspace}

\newcommand{\unit}[1]{\ensuremath{\mathrm{#1}}\xspace}

\newcommand{\cm}{\,\unit{cm}}

\newcommand{\sfluxunit}{\,\ensuremath{\frac{1}{\unit{cm}^2\unit{s}}}}

\newcommand{\Xsunit}{\,\ensuremath{\unit{\frac{1}{cm}}}}
\newcommand{\sourceunit}{\,\ensuremath{\unit{\frac{\mathrm{n}}{s}}}}

\hypersetup{
    pdftitle={Nonlinear Diffusion Acceleration of the Least-Squares Transport Equation in Geometries with Voids},
    pdfauthor={Hans R Hammer},
    pdfsubject={Nonlinear diffusion acceleration for Least squares in voids},
    pdfkeywords={Void, Least Square, NDA},
    colorlinks=true,
    linkcolor=black,
    citecolor=black,
    filecolor=black,
    urlcolor=black,
}

\renewcommand{\drift}[1][]{\ensuremath{\hat{\alpha}\addgroup{#1}}\xspace}

\title{Nonlinear Diffusion Acceleration of the Least-Squares Transport Equation in Geometries with Voids}
\author[1,2]{Hans R. Hammer\thanks{email: hrhammer@lanl.gov}\xspace}
\author[2]{Jim E. Morel\thanks{email: morel@tamu.edu}}
\author[3]{Yaqi Wang}
\affil[1]{Los Alamos National Laboratory - T3: Fluid Dynamics and Solid Mechanics\\ Bikini Atoll Road, Los Alamos, NM, 87545}
\affil[2]{Texas A\&M University - Department of Nuclear Engineering\\ 3133 TAMU, College Station, TX 77843-3133}
\affil[3]{Idaho National Laboratory, 1955 N. Fremont Ave, Idaho Falls, ID 83415}
\date{                               
    \vspace{40mm}
    Number of pages: \pageref*{LastPage} \\
    Number of tables: \totaltables \\
    Number of figures: \totalfigures
}

\begin{document}
    
    \clearpage\maketitle
    \thispagestyle{empty}
    \pagebreak
    
    \begin{abstract}
        In this paper we show the extension of the Nonlinear-Diffusion Acceleration (NDA) to geometries containing small voids using a weighted least-squares (WLS) high order equation. Even though the WLS equation is well defined in voids, the low-order drift diffusion equation was not defined in materials with a zero cross section.
        
        This paper derives the necessary modifications to the NDA algorithm. We show that a small change to the NDA closure term and a non-local definition of the diffusion coefficient solve the problems for voids regions. These changes do not affect the algorithm for optical thick material regions, while making the algorithm well defined in optically thin ones. We use a Fourier analysis to perform an iterative analysis to confirm that the modifications result in a stable and efficient algorithm. 
        
        Numerical results of our method will be presented in the second part of the paper. We test this formulation with a small, one-dimensional test problem. Additionally we present results for a modified version of the C5G7 benchmark containing voids as a more complex, reactor like problem. We compared our results to PDT, Texas A\&M's transport code, utilizing a first order discontinuous formulation as reference and the self-adjoint angular flux equation with void treatment (\saaft), a different second order form. The results indicate, that the NDA WLS performed comparably or slightly worse then the asymmetric \saaft, while maintaining a symmetric discretization matrix.\par
                
        \vspace{1em}\noindent\textbf{Keywords} --- Neutron transport, Weighted Least-Squares, Nonlinear Diffusion Acceleration, Voids, Fourier analysis
    \end{abstract}
    
    \vfill
    \pagebreak

\section{Introduction}
    
    Nonlinear Diffusion Acceleration (NDA) is an effective acceleration technique for the \sn source iteration method~\cite{adams_fast_2002} in optically thick media. The NDA method is especially of interest for reactor physics problems, since it is easily adopted to solve criticality problems ~\cite{park_nonlinear_2012}. Additionally, NDA enforces conservation of particles for the weighted least-squares (WLS) equation~\cite{peterson_conservative_2015,hammer_weighted_2018} and therefore is an important improvement for the WLS equation even in non-diffusive cases.
    
    Even though the WLS high-order equation is well defined in voids, the low-order drift diffusion equation is unbounded for both DSA and NDA. This is caused by the standard diffusion coefficient, having the total cross section in the denominator. Additionally, the standard way to evaluate the drift vector for the NDA algorithm leads to another division by zero in voids. \par
    
    The purpose of this paper is to develop a conservative void-compatible NDA scheme for our WLS transport formulation. While the previous paper \cite{hammer_weighted_2018} focused on the WLS high-order equation of our High-Order Low-order (HOLO) system, this paper now describes the extension of the NDA low-order equation for geometries with small voids. The proposed solution includes the use of a non-local diffusion coefficient~\cite{morel_non-local_2007} and a combination of current formulations for the drift-vector. The latter is a correction term of the diffusion equation, which is informed by the full angular flux solution. We present numerical results for test problems containing voids. The test problems are Reed's problem, a two region, one-dimensional problem and a modified version of the C5G7 benchmark~\cite{smith_benchmark_2004} with a more complex, reactor like geometry.
    
    In this paper we will perform an iterative analysis with a Fourier analysis and numerical results. The structure of this paper is as follows: we first introduce the WLS form of the transport equation and the corresponding NDA scheme~\cite{peterson_conservative_2015,hammer_weighted_2018}. In the following chapter we perform a multi-region Fourier analysis with a general form of the NDA drift vector. This allows us to study the effect of different closure terms and of the non-local diffusion coefficient on the convergence rate to ensure a stable and efficient algorithm. We compare our analytical results to a numerical Fourier analysis to confirm our findings. In \cref{ch:results} we present one dimensional results and analyze the behavior of WLS NDA in void regions. We use a modified version of the C5G7 benchmark to test our method on a more complex, reactor like problem. Finally, in the last section we summarize our findings and draw conclusions.\par
    
\section{Theory} \label{ch:second_order}

\subsection{Nonlinear Diffusion Acceleration for WLS} \label{sec:wls}

    The mono-energetic weighted least-squares (WLS) equation, which we derived previously~\cite{hammer_weighted_2018}, can be written as
    \begin{subequations} \label{eq:wls_equation}
        \begin{multline} \label{eq:wls_transport}
            - \direction \cdot \del \left[ \weight \direction \cdot \del \psi\right]
            - \direction \cdot \psi \del \left[ \weight \sigt\right]
            + \weight \sigt^2\psi \\
            = -\direction\cdot\del\left[\weight \sum_{l=0}^{L}\sum_{p=-l}^{l} \frac{2l+1}{4\pi} \spharmonic\left(\direction\right) \sigl{l} \mflux
            + \weight\frac{\bar{\nu}\sigf}{4\pi} \phi
            + \weight\frac{q}{4\pi}\right] \\
            + \weight\sigt\sum_{l=0}^{L}\sum_{p=-l}^{l} \frac{2l+1}{4\pi} \spharmonic\left(\direction\right) \sigl{l} \mflux
            + \weight\frac{\sigt \bar{\nu}\sigf}{4\pi}\phi
            + \weight\frac{q\sigt}{4\pi}.
        \end{multline}
        where 
        \begin{equation} \label{eq:def_void_weight}
        \weight \equiv \min\left(\frac{1}{\sigt},\,\weight[_\mathrm{max}]\right),
        \end{equation}
        denotes the weight function with \sigt the total cross section. The corresponding boundary conditions on the domain boundary \boundary are
        \begin{align} \label{eq:wls_strong_bc}
            \psi\left(\vec{x}_b, \direction\right) &= \psi^\mathrm{inc}\left(\vec{x}_b, \direction \right), \qquad \forall\vec{x}_b \in \boundary, \quad \direction \cdot \normal < 0  \\
            \direction \cdot \del \psi\left(\vec{x}_b\right) + \sigt \psi\left(\vec{x}_b\right) &= \sum_{l=0}^{L}\sum_{p=-l}^{l} \frac{2l+1}{4\pi} \spharmonic\left(\direction\right) \sigl{l} \mflux
            + \frac{\bar{\nu}\sigf}{4\pi} \phi
            + \frac{q}{4\pi}
            \qquad \direction \cdot \normal > 0.
        \end{align}
    \end{subequations}
    \(\psi(\vec{x},\direction)\) is the angular flux with \(\vec{x} \in \domain\), \(\direction\in 4\pi\) (\(4\pi\) represents the entire unit sphere), \(\spharmonic\left(\direction\right)\) are the spherical harmonics with \(\sigl{l}\) the scattering cross section moments and \(q\) is the distributed source.

    The left-hand side of this equation is self-adjoint and decouple for all directions, which makes it compatible with source iterations. \(\weight_\mathrm{max} \) denotes a maximum value for the weight function. This definition will make the WLS equation well defined in voids and maintain the symmetric positive-definite properties of the resulting discretized matrix.

    The resulting mono-energetic WLS weak form used in this paper is defined as follows:
    Given a trial space \(W_\domain\), consisting of continuous basis functions and an angular quadrature \(\left\{\direction_m, \omega_m\right\}_{m=1}^{M}\), the weak form for a specific direction is as follows:
    Find \(\atestfct\in W_\domain\) such that
    \begin{multline}\label{eq:wls_weak_void}
        \left(\weight\direction\cdot\del\psi, \direction\cdot\del\atestfct + \sigt\atestfct\right)_\domain
        + \left(\weight\sigt\psi, \direction\cdot\del\atestfct + \sigt\atestfct\right)_\domain \\
        + \left<\weight f \left(\psi - \psi^\mathrm{inc}\right), \atestfct\left| \direction \cdot \vec{n}\right|\right>_{\boundary^-}
        = \left(\weight\sum_{l=0}^{L}\sum_{p=-l}^{l}\frac{2l+1}{4\pi} \spharmonic\left(\direction\right) \sigl{l} \mflux, \direction\cdot\del\atestfct + \sigt\atestfct\right)_\domain \\
        + \left(\weight\frac{\nu\sigf}{4\pi}\phi,  \direction\cdot\del\atestfct + \sigt\atestfct\right)_\domain
        + \left(\weight\frac{q}{4\pi}, \direction\cdot\del\atestfct + \sigt\atestfct\right)_\domain
    \end{multline}
    where 
    \begin{equation}
    \left(\vphantom{\del}\cdot, \cdot\right)_\domain \equiv \int_\domain \dx[V]
    \end{equation} 
    is the standard spatial inner product and 
    \begin{equation}
    \left<\vphantom{\del}\cdot, \cdot\right>_\boundary \equiv \oint_\boundary \dx[A]
    \end{equation} 
    is the corresponding surface integral. We chose to use the optional weak boundary condition over \({\boundary^-}\) the portion of the boundary for which \(\direction \cdot \normal < 0\). The boundary weight function is defined as
    \begin{equation} \label{eq:ls_bc_7}
        f \equiv \max\left(\sigt, \frac{1}{h}\right),
    \end{equation}
    where \(h\) denotes a characteristic length constant of the boundary cell. \par

	We use an inconsistent, but conservative form of the NDA, which enforces conservation for the whole system~\cite{hammer_weighted_2018}. The NDA drift-diffusion equation is
    \begin{equation} \label{eq:drift_diffusion}
    - \del \cdot \left[\DC \del \phi\right] - \del \cdot \left[\drift \phi\right] + \siga \phi = q.
    \end{equation}
    where the drift vector
    \begin{equation} \label{eq:drift_vector}
        \drift \equiv \frac{1}{\phi} \left(\frac{1}{\sigt} \sum_{m=1}^{M}\omega_m\direction_m \left(\direction_m \cdot \del \psi_m \right) - \DC \del \phi \right)
    \end{equation}
    is an additive correction to Fick's law~\cite{smith_nodal_1983} and the diffusion coefficient is defined as
    \begin{equation} \label{eq:diffusion_coefficient}
        \DC \equiv \frac{1}{3\sigt}.
    \end{equation}
    
    Multiplying \cref{eq:drift_diffusion} by a test function \(\phi^*\) and integrating over the domain gives the corresponding weak form. 
    Applying integration by parts on the current term gives
    \begin{equation} \label{eq:low_order_eq}
        \left(\DC \del \phi, \del\testfct \right)_\domain
        +\left(\drift \phi, \del\testfct \right)_\domain 
        + \left< \frac{1}{4}\kappa \phi - J^\mathrm{\,in}, \testfct \right>_{\boundary}
        + \left(\siga \phi, \testfct \right)_\domain
        =\left(q, \testfct \right)_\domain.
    \end{equation}
    with the vacuum boundary coefficient as
    \begin{equation} \label{eq:kappa}
        \kappa \equiv \frac{4}{\phi} \sum_{\normal \cdot \direction_m > 0} \omega_m \left|\normal \cdot \direction_m\right| \psi_m
    \end{equation}

\subsection{Multi-region Fourier Analysis for NDA} \label{sec:nda_fa_homo}

    In this section we derive the tool to analyze a general form of the discretized NDA WLS equations. A measure of the efficiency of an iteration scheme is the error reduction per iteration. For the analysis of the convergence behavior we consider the isotropic case in an infinite medium. Given the exact scalar flux solution \(\phi\) the error is obtained by
    \begin{equation} \label{eq:si_error}
        \error{\phi}\iter{k} \equiv {\phi\iter{k} - \phi}.
    \end{equation}
    where \(\error{\phi}\iter{k}\) denotes the scalar flux error at the \(k\)th iteration. The error reduction can be expressed asymptotically with sufficient large \(k\) as
    \begin{equation} \label{eq:si_rho}
        \rho \equiv \frac{\error{\phi}\iter{k+1}}{\error{\phi}\iter{k}}
    \end{equation}
    with \(\rho\) the spectral radius. In practice this is seen after a reasonable number of iterations. \par
    
    For the infinite homogeneous problem with the total cross section \sigt and the isotropic scattering cross section \sigs, the spectral radius of the unaccelerated source iteration scheme is \(\rho = c\)~\cite{adams_fast_2002}, where \(c\) denotes the scattering ratio
    \begin{equation} \label{eq:scattering_ratio}
        c \equiv \frac{\sigs}{\sigt}.
    \end{equation}
    Hence for highly diffusive media with \(\sigs \approx \sigt \), the convergence rate of source iterations is not sufficient for an efficient use in computational applications. \par

\begin{figure}[h]
        \includegraphics[width=0.8\textwidth]{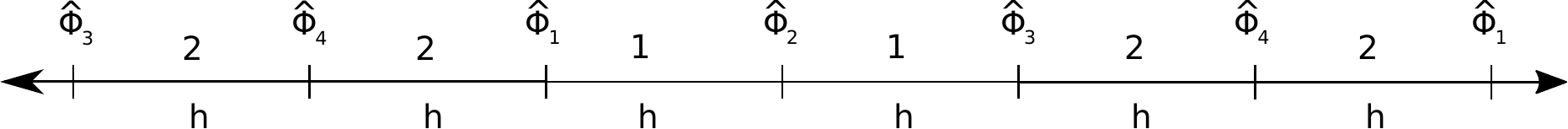}
        \caption{Section of an infinite mesh for the Fourier analysis with 2 regions and 4 periodic cells.}
        \label{fig:fa_mesh}
    \end{figure}

    To investigate the convergence properties of a modified scheme we need to be able to handle multiple material regions, including void regions. Therefore we derive a numerical Fourier analysis with multiple regions. This tool uses a Fourier transformation to determine the spectral radius for an iterative technique. Since the discretization can change the spectral radius, especially for a independently differenced acceleration formulation \cite{adams_fast_2002}, we study the one-dimensional discretized equations. For this we assume an infinite periodic mesh with no assumptions regarding the periodicity of the solution. \Cref{fig:fa_mesh} shows a mesh for a problem with two material regions and 2 cells per region. The mesh extends infinitely beyond the section shown, repeating the same structure. Nevertheless we do not assume the solution is periodic. The solution has the form of a vector, i.e. four variable types, multiplied by a complex exponential depending on the position on the whole, infinite mesh. The exponential allows the solution to be non-periodic. \par
    
\subsubsection{High-order equation} \label{sec:fa_wls}
    First we consider the high order WLS equation to be able to derive the NDA closure terms. The one-dimensional, isotropic equation for WLS, \cref{eq:wls_equation}, in the weak form at iteration index \(k\) is~\cite{hammer_weighted_2018}
    \begin{multline} \label{eq:fa_wls_infinite}
        \left(\weight\mu\ddx\psi\iter{k+\half}, \mu\ddx\atestfct + \sigt\atestfct\right)_\domain
        + \left(\weight\sigt\psi\iter{k + \half}, \mu\ddx\atestfct + \sigt\atestfct\right)_\domain \\
        = \left(\weight\frac{c\sigt}{2}\phi\iter{k}, \mu\ddx\atestfct + \sigt\atestfct\right)_\domain
        + \left(\weight\frac{q}{2}, \mu\ddx\atestfct + \sigt\atestfct\right)_\domain
    \end{multline}
    with the scattering cross section
    \begin{equation}
        \sigs = c \sigt.
    \end{equation}
    Using the problem's exact solution \(\psi\) with
    \begin{equation} \label{eq:fa_exact_psi}
        \psi = \psi\iter{k} + \error{\psi}\iter{k}
    \end{equation}
    where \(\error{\psi}\iter{k}\) is the error in the \(k\)th iteration and subtracting it from \cref{eq:fa_wls_infinite} gives the WLS equation for the error
    \begin{multline} \label{eq:fa_wls_error}
        \left(\weight\mu\ddx\error{\psi}\iter{k + \half}, \mu\ddx\atestfct\right)_\domain
        + \left(\weight\mu\ddx\error{\psi}\iter{k + \half}, \sigt\atestfct\right)_\domain
        + \left(\weight\sigt\error{\psi}\iter{k + \half}, \mu\ddx\atestfct\right)_\domain \\
        + \left(\vphantom{\ddx}\weight\sigt\error{\psi}\iter{k + \half}, \sigt\atestfct\right)_\domain
        = \left(\weight\frac{c\sigt}{2}\error{\phi}\iter{k}, \mu\ddx\atestfct\right)_\domain
        + \left(\weight\frac{c\sigt}{2}\error{\phi}\iter{k}, \sigt\atestfct\right)_\domain.
    \end{multline} \par

    We apply first order continuous finite elements with the cell index \(i\), where integer indices indicate the interior of a cell, while half indices denote the mesh vertices, and obtain
    \begin{multline} \label{eq:fa2_wls_discretized}
        \left(\mu^2\left(\frac{\weight[i]}{h_{i}} + \frac{\weight[i+1]}{h_{i+1}}\right) + \mu\left(\weight[i]\sigt[i] - \weight[i+1]\sigt[i+1]\right) + \frac{\weight[i]\sigt[i]^2h_{i}}{3} + \frac{\weight[i+1]\sigt[i+1]^2h_{i+1}}{3}\right)\psi_{i+\half}\iter{k+\half}  \\
        + \left(-\frac{\mu^2\weight[i]}{h_{i}} + \frac{\weight[i]\sigt[i]^2h_{i}}{6}\right)\psi_{1-\half}\iter{k+\half}
        + \left(-\frac{\mu^2\weight[i+1]}{h_{i+1}}+ \frac{\weight[i+1]\sigt[i+1]^2h_{i+1}}{6}\right)\psi_{1+\half[3]}\iter{k+\half} \\
        = \left(\mu\frac{\weight[i]c_{i}\sigt[i]}{4} + \frac{\weight[i]c_{i}\sigt[i]^2h_{i}}{12}\right)\phi_{i-\half}\iter{k}
        + \left(-\mu\frac{\weight[i+1]c_{i+1}\sigt[i+1]}{4} + \frac{\weight[i+1]c_{i+1}\sigt[i+1]^2h_{i+1}}{12}\right)\phi_{i+\half[3]}\iter{k} \\
        + \left(\mu\frac{\weight[i]c_{i}\sigt[i]-\weight[i+1]c_{i+1}\sigt[i+1]}{4} + \frac{\weight[i]c_{i}\sigt[i]^2h_{i}}{6} + \frac{\weight[i+1]c_{i+1}\sigt[i+1]^2h_{i+1}}{6}\right)\phi_{i+\half}\iter{k}.
    \end{multline}
    Based on the Fourier ansatz
    \begin{equation} \label{eq:fa_transformation}
        \error{\psi}\left(x, \mu\right) = \int_0^\infty \hat{\psi}\left(\lambda, \mu\right) \e{\img \lambda \sigt x}\dx[\lambda]
    \end{equation}
    we use the following discrete ansatz
    \begin{equation} \label{eq:fa2_def_angular_flux}
        \psi_{i+\half}\iter{k} = \int_0^{\lambda_{\max}} \hat{\psi}\iter{k}_{i+\half} \left(\lambda, \mu\right) \e{\img\lambda x_{i+\half}} \dx[\lambda]
    \end{equation}
    where \(\lambda_{\max}\) is the maximal frequency supported by the mesh. The corresponding flux moments are
    \begin{align} \label{eq:fa2_def_scalar_flux}
        \hat{\phi}_{i+\half}\iter{k}\left(\lambda\right) &= \intpolar \hat{\psi}_{i+\half}\iter{k} \left(\lambda, \mu\right) \dmu \\
        \label{eq:fa2_def_current}
        \hat{J}_{i+\half}\iter{k}\left(\lambda\right) &= \intpolar \mu\hat{\psi}_{i+\half}\iter{k} \left(\lambda, \mu\right) \dmu \\
        \label{eq:fa2_def_pressure}
        \hat{\xi}_{i+\half}\iter{k}\left(\lambda\right) &= \intpolar \mu^2\hat{\psi}_{i+\half}\iter{k} \left(\lambda, \mu\right) \dmu.
    \end{align}
    The periodic geometry gives
    \begin{equation}
        \hat{\psi}^{k}_{i+\half}\left(\lambda, \mu\right) = \hat{\psi}^{k}_{i+\half+N}\left(\lambda, \mu\right)
    \end{equation}
    where \(N\) denotes the number of cells after which the geometry is repeated. As mentioned above, this condition does not require the solution for a specific frequency \(\lambda\) to be periodic within \(N\) cells due to the complex exponential.
    Substituting the ansatz into \cref{eq:fa2_wls_discretized} gives an equation for a specific frequency \(\lambda\). Using the definitions in \cref{eq:fa2_def_angular_flux}, \cref{eq:fa2_wls_discretized} can be written as matrix equation
    \begin{equation} \label{eq:fa2_wls_matrix}
        \mat{A}\vec{\hat{\psi}}\iter{k+\half} = \mat{B}\vec{\hat{\phi}}\iter{k}
    \end{equation}
    with the solution
    \begin{equation} \label{eq:fa2_psi}
        \vec{\hat{\psi}}\iter{k+\half} = \inv{\mat{A}}\mat{B}\vec{\hat{\phi}}\iter{k}
    \end{equation}
    where \mat{A} is the streaming and collision matrix with the entries for \(i=1\dots N,\:m=1\dots M\) (index \(s_i = m\cdot N + \left(i \mod N\right)\))
    \begin{subequations}
        \begin{align}
            a_{s_{i}s_{i-1}} &= \weight[i]\left(-\frac{\mu_m^2}{h_{i}} + \frac{\sigt[i]^2h_{i}}{6}\right)\e{-\img\lambda h_i} \\
            a_{s_{i}s_{i}} &= \weight[i]\left(\frac{\mu_m^2}{h_{i}} + \sigt[i]\mu_m + \frac{\sigt[i]^2 h_{i}}{3}\right)
            + \weight[i+1]\left(\frac{\mu_m^2}{h_{i+1}} - \mu_m\sigt[i+1] + \frac{\sigt[i+1]^2h_{i+1}}{3}\right) \\
            a_{s_{i}s_{i+1}} &= \weight[i+1]\left(-\frac{\mu_m^2}{h_{i+1}}+ \frac{\sigt[i+1]^2h_{i+1}}{6}\right)\e{\img\lambda h_{i+1}}
        \end{align}
    \end{subequations}
    and \mat{B} the scattering matrix for \(i=1\dots N,\:m=1\dots M\)
    (with the indices \(s_i = m\cdot I + i\) and \(t_i = \left(i \mod N\right)\))
    \begin{subequations}
        \begin{align}
            b_{s_{i}t_{i-1}} &= \weight[i]\left(\mu_m\frac{c_{i}\sigt[i]}{4} + \frac{c_{i}\sigt[i]^2h_{i}}{12}\right)\e{-\img\lambda h_{i}} \\
            b_{s_{i}t_{i}} &= \weight[i]c_{i}\sigt[i]\left(\frac{\mu_m}{4} + \frac{\sigt[i]h_{i}}{6}\right)
            + \weight[i+1]c_{i+1}\sigt[i+1]\left(-\frac{\mu_m}{4} + \frac{\sigt[i+1]h_{i+1}}{6}\right) \\
            b_{s_{i}t_{i+1}} &= \weight[i+1]\left(-\mu_m\frac{c_{i+1}\sigt[i+1]}{4} + \frac{c_{i+1}\sigt[i+1]^2h_{i+1}}{12}\right)\e{\img\lambda h_{i+1}}.
        \end{align}
    \end{subequations}    
    The \sn angular discretization gives for the scalar flux \cref{eq:fa2_def_scalar_flux}
    \begin{equation}
        \hat{\phi}^{k}_{i+\half} = \sum_{m=1}^{M} \omega_m \hat{\psi}^{k}_{i+\half, m}
    \end{equation}
    which is written in vector form
    \begin{align} \label{eq:fa2_quad_phi}
        \vec{\hat{\phi}}^{k+\half} &= \mat{W_0}\vec{\hat{\psi}}^{k+\half} \notag\\
                             &= \mat{W_0}\inv{\mat{A}}\mat{B}\vec{\hat{\phi}}\iter{k}
    \end{align}
    where \mat{W_0} is the zeroth moment angular quadrature matrix. Accordingly, the neutron current \cref{eq:fa2_def_current} is
    \begin{align} \label{eq:fa2_quad_j}
        \vec{\hat{J}}^{k+\half} &= \mat{W_1}\vec{\hat{\psi}}^{k+\half} \notag\\
                          &= \mat{W_1}\inv{\mat{A}}\mat{B}\vec{\hat{\phi}}\iter{k}
    \end{align}
    and the second moment \cref{eq:fa2_def_pressure} is
    \begin{align} \label{eq:fa2_quad_xi}
        \vec{\hat{\xi}}^{k+\half} &= \mat{W_2}\vec{\hat{\psi}}^{k+\half}\notag\\
                            &= \mat{W_2}\inv{\mat{A}}\mat{B}\vec{\hat{\phi}}\iter{k}
    \end{align}
    with \mat{W_1} and \mat{W_2} the first and second moment angular quadrature matrices, respectively. \par
    
    The spectral radius for source iterations with WLS transport is the absolute value of the eigenvalue with the largest magnitude of the systems matrix in \cref{eq:fa2_quad_phi} with \(\phi\iter{k+1} = \phi\iter{k+\half}\). It can be easily found using numerical libraries such as SciPy \cite{jones_scipy.org_2001}. The results confirm a spectral radius of \(c\) for standard source iterations. \par

\subsubsection{Low order equation}
    Using the derivation of the NDA the closure term of the drift-diffusion equation contains the Eddington form of the neutron current (\cref{eq:drift_vector}). This is problematic in voids because of the total cross section in the denominator. The consistent NDA derivation for the \saaft equation \cite{wang_diffusion_2014} contains the Eddington form and the direct form of the neutron current, weighted by a function \(\tau\left(\sigt\right)\)
    \begin{equation} \label{eq:saaf_nda_drift}
        \drift\iter{k+\half} = \frac{1}{\phi\iter{k+\half}} \left(\tau \sum_{m=1}^{M} \direction_m \left( \direction_m \cdot \del \psi_m\iter{k+\half} \right) 
        \vphantom{\sum_{m=1}^{M}} -\bracket{\parenthesis{1 - \tau \sigt} \current\,\iter{k+\half}} - \DC \del \phi\iter{k+\half}\right).
    \end{equation}
    
    Both expressions can be expressed using a general, linearized drift vector
    \begin{equation} \label{eq:fa2_drift_vector}
        \drift\iter{k+\half}\phi\iter{k+1} = \frac{p}{\sigt}\xi\iter{k+\half} - \tilde{p}J\iter{k+\half} - \DC\ddx \phi\iter{k+\half}
    \end{equation}
    where \(p\) and \(\tilde{p}\) are weights depending on the specific formulation of the drift closure. With this, the general, linearized low-order error equation becomes
    \begin{multline} \label{eq:fa2_nda_error}
        \left(\DC\ddx\error{\phi}\iter{k+1}, \ddx\testfct\right)_\domain
        + \left(\left(1-c\right)\sigt\error{\phi}\iter{k+1}, \testfct\right)_\domain \\
        = -\left(\frac{p}{\sigt}\ddx\error{\xi}\iter{k+\half}, \ddx\testfct\right)_\domain
        + \left(\tilde{p}\error{J}\iter{k+\half}, \ddx\testfct\right)_\domain
        + \left(\DC\ddx\error{\phi}\iter{k+\half}, \ddx\testfct\right)_\domain
    \end{multline}
    with the finite-element discretization
    \begin{multline}
        \left(\frac{\DC_{i}}{h_{i}} + \frac{\DC_{i+1}}{h_{i+1}} + \frac{\left(1-c_{i}\right)\sigt[i]h_{i}}{3} + \frac{\left(1-c_{i+1}\right)\sigt[i+1]h_{i+1}}{3}\right)\phi\iter{k+1}_{i+\half}  \\
        + \left( - \frac{D_{i}}{h_{i}} + \frac{\left(1 - c_{i}\right)\sigt[i]h_{i}}{6}  \right)\phi\iter{k+1}_{i-\half}
        + \left( - \frac{D_{i+1}}{h_{i+1}} + \frac{\left(1 - c_{i+1}\right)\sigt[i+1]h_{i+1}}{6}\right)\phi\iter{k+1}_{i+\half[3]}  \\
        = -\left(\left(\frac{p_{i}}{\sigt[i]h_{i}} + \frac{p_{i+1}}{\sigt[i+1] h_{i}}\right)\xi\iter{k+\half}_{i+\half} - \frac{p_{i}}{\sigt[i] h_{i}}\xi\iter{k+\half}_{i-\half} - \frac{p_{i+1}}{\sigt[i+1] h_{i+1}}\xi\iter{k+\half}_{i+\half[3]}\right)
        + \frac{\tilde{p}_i}{2}\left(\error{J}\iter{k+\half}_{i-\half} + \error{J}\iter{k+\half}_{i+\half}\right)  \\
        - \frac{\tilde{p}_{i+1}}{2}\left(\error{J}\iter{k+\half}_{i+\half} + \error{J}\iter{k+\half}_{i+\half[3]}\right)
        + \left(\left(\frac{\DC_{i}}{h_{i}} + \frac{\DC_{i+1}}{h_{i+1}}\right)\phi\iter{k+\half}_{i+\half} - \frac{\DC_{i}}{h_{i}} \phi\iter{k+\half}_{i-\half} - \frac{\DC_{i+1}}{h_{i+1}} \phi\iter{k+\half}_{i+\half[3]}\right).
    \end{multline}
    Using the Fourier ansatz \cref{eq:fa2_def_scalar_flux,eq:fa2_def_current,eq:fa2_def_pressure} we can write this as vector equation
    \begin{equation} \label{eq:fa_nda_phi}
    \mat{C}\vec{\hat{\phi}}\iter{k+1} = \mat{E}\vec{\hat{\phi}}\iter{k+\half} + \mat{F}\left(\tilde{p}\right)\vec{\hat{J}}\iter{k+\half} + \mat{G}\left(p\right)\vec{\hat{\xi}}\iter{k+\half} 
    \end{equation}
    with the entries of the diffusion matrix \(\mat{C}\) for \(i = 1\dots N\)
    \begin{subequations}
        \begin{align}
            c_{ii-1} &= \left( - \frac{D_{i}}{h_{i}} + \frac{\left(1 - c_{i}\right)\sigt[i]h_{i}}{6}  \right)\e{-\img\lambda h_{i}}\\
            c_{ii} &= \frac{\DC_{i}}{h_{i}} + \frac{\DC_{i+1}}{h_{i+1}} + \frac{\left(1-c_{i}\right)\sigt[i]h_{i}}{3} + \frac{\left(1-c_{i+1}\right)\sigt[i+1]h_{i+1}}{3} \\
            c_{ii+1} &= \left( - \frac{D_{i+1}}{h_{i+1}} + \frac{\left(1 - c_{i+1}\right)\sigt[i+1]h_{i+1}}{6}\right)\e{\img\lambda h_{i+1}},
        \end{align}
    \end{subequations}
    for the zeroth moment drift matrix \(\mat{E}\)
    \begin{subequations}
        \begin{align}
            e_{ii-1} &= -\frac{\DC_{i}}{h_{i}}\e{-\img\lambda  h_{i}} \\
            e_{ii} &= \frac{\DC_{i}}{h_{i}} + \frac{\DC_{i+1}}{h_{i+1}} \\
            e_{ii+1} &= -\frac{\DC_{i+1}}{h_{i+1}} \e{\img\lambda h_{i+1}}
        \end{align}
    \end{subequations}
    and for the first moment drift matrix \(\mat{F}\)
    \begin{subequations} \label{eg:fa2_current}
        \begin{align}
            f_{ii-1} &= \frac{\tilde{p}_{i}}{2}\e{-\img\lambda  h_{i}} \\
            f_{ii} &=  \frac{\tilde{p}_{i}}{2} - \frac{\tilde{p}_{i+1}}{2} \\
            f_{ii+1} &=  -\frac{\tilde{p}_{i+1}}{2}\e{\img\lambda  h_{i+1}}.
        \end{align}
    \end{subequations}
    Finally we get for the second moment drift matrix \(\mat{G}\)
    \begin{subequations} \label{eq:fa2_eddington}
        \begin{align}
            g_{ii-1} &= \frac{p_{i}}{\sigt[i] h_{i}}\e{-\img\lambda  h_{i}} \\
            g_{ii} &= -\frac{p_{i}}{\sigt[i]h_{i}} - \frac{p_{i+1}}{\sigt[i+1] h_{i+1}} \\
            g_{ii+1} &=  \frac{p_{i+1}}{\sigt[i+1] h_{i+1}}\e{\img\lambda h_{i+1}}.
        \end{align}
    \end{subequations}
    Substituting \cref{eq:fa2_quad_phi,eq:fa2_quad_j,eq:fa2_quad_xi} into \cref{eq:fa_nda_phi} and with the angular flux solution in \cref{eq:fa2_psi} gives
    \begin{equation} \label{eq:fa2_matrix_rho}
        \vec{\hat{\phi}}\iter{k+1} 
        = \inv{\mat{C}}\big(\mat{E}\mat{W_0}
        + \mat{F}\left(\tilde{p}\right)\mat{W_1} 
        + \mat{G}\left(p\right)\mat{W_2}\big)\inv{\mat{A}}\mat{B}\vec{\hat{\phi}}\iter{k}
    \end{equation}
    The spectral radius of the acceleration scheme can be found as the absolute value of the eigenvalue with the largest magnitude of the system matrix in \cref{eq:fa2_matrix_rho}. \par
    
\subsection{Drift vector formulations}

    \begin{table}[t]
        \caption{Weight factors for the different current formulations for the general drift vector.}
        \setlength{\tabcolsep}{12pt}
        \begin{tabular}{lrr}
            \toprule
            Formulation &          \(p\) &      \(\tilde{p}\) \\ \midrule
            Eddington   &              1 &                  0 \\
            Current     &              0 &                  1 \\
            Combined    & \(\hat{\tau}\) & \(1 - \hat{\tau}\) \\
            \(\tau\)    & \(\sigt\tau \) &  \(1 - \sigt\tau\) \\ \bottomrule
        \end{tabular}
        \setlength{\tabcolsep}{6pt}
        \label{tab:general_drift_weight}
    \end{table}
    
    All the formulation for the general, linearized drift vector that we considered can be found in \cref{tab:general_drift_weight}. The weight function for the combined formulation is defined as
    \begin{equation}
        \hat{\tau} \equiv \begin{cases}
            1,\qquad \sigt h \ge \hat{\zeta} \\
            0,\qquad \sigt h < \hat{\zeta} \\
        \end{cases}
    \end{equation}
    with \(\hat{\zeta} = \tento{-2}\) and for the \(\tau\)-formulation based on the NDA for \saaft~\cite{wang_diffusion_2014} as
    \begin{align} \label{eq:saaf_tau}
        \tau \equiv \begin{cases}
            \frac{1}{\sigt}, & \sigt h \ge \zeta \\
            \frac{h}{\zeta}, & \sigt h < \zeta
        \end{cases}
    \end{align}
    with \(\zeta = 0.5\). \(\hat{\zeta}\)  and \(\zeta\) are parameters of the formulations with their default values. \par
    
\begin{figure}[tp]
        \begin{minipage}{0.8\textwidth}
            \setlength{\figheight}{6cm}
            \begin{tikzpicture}
	\begin{semilogxaxis}[
            xmin = 1e-3, xmax = 1e3,
            ymin = 0, ymax = 1.1,
            ytick = {0,0.25,...,1.0},
            y tick label style={/pgf/number format/.cd, scaled y ticks = false, fixed},
            x tick label style={/pgf/number format/.cd, scaled x ticks = false, fixed},,
            height=\figheight,
            width=\textwidth,
            grid=major,
            legend style={at={(0.05, 0.95)},anchor=north west},
            legend columns = 1,
            cycle list name = color list,
            xlabel = {Optical Thickness \(\sigt h \) [-]},
            ylabel = {Spectral radius \(\rho\) [-]}
		]

		\addplot [red, thick, mark = *, mark repeat = 8, mark phase = 5] table [x = x, y = max] {data/fourier_analysis/two_region_spectral_lambda_nda_single_cell_pure_scatterer.tab};
		\addlegendentry{Eddington}

		\addplot [teal, thick, mark = square*, mark repeat = 8] table [x = x, y = max] {data/fourier_analysis/two_region_spectral_lambda_j_single_cell_pure_scatterer.tab};
		\addlegendentry{Current}

        \addplot [cyan, thick, mark = triangle*, mark repeat = 8] table [x = mfp, y = rho] {data/fourier_analysis/two_region_spectral_lambda_cvoid_single_cell_pure_scatterer.tab};
        \addlegendentry{Combined}

	\end{semilogxaxis}
\end{tikzpicture}
        \end{minipage}
        \caption{Spectral radius for \(c=1\) as function of the optical cell thickness for the Eddington, Current and Combined NDA formulations in an infinite homogeneous material.}
        \label{fig:fourier_analysis}
    \end{figure}
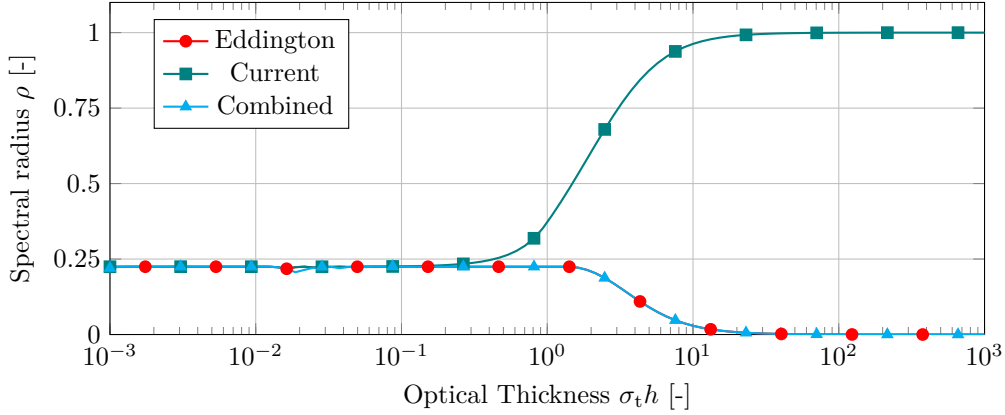

    First we compared these formulations in the case of an infinite, homogeneous medium with \(c = 1\). The spectral radii for the Eddington, Current and Combined formulation are shown in \cref{fig:fourier_analysis}. The results indicate clearly, that the Current formulation has a loss of efficiency for optical thick cells, while the Eddington formulation actually increases the convergence for optical thick cells. In the optical thin regime, these two formulations show the same spectral radius. The difference between the two formulations arise from the discretization of the drift-vector. \Cref{eg:fa2_current} will result for a homogeneous medium in a two point derivative, skipping the center node, whereas \cref{eq:fa2_eddington} gives a three point discretization. The two point scheme is known to be problematic for high frequency modes, resulting in a decrease of convergence. From these observations we derived the Combined formulation, combining the advantages of the other two formulations. These advantages are the better convergence of the Eddington form for optical thick cells and the better conditioning for voids and near voids of the Current form. The spectral radius of the Combined formulation is also shown in \cref{fig:fourier_analysis}. The parameter \(\hat{\zeta} = \tento{-2}\) was chosen, so the switch between the two formulations is sufficient away the increase of the Current formulation, but without the Eddington formulation being already ill-conditioned.
    
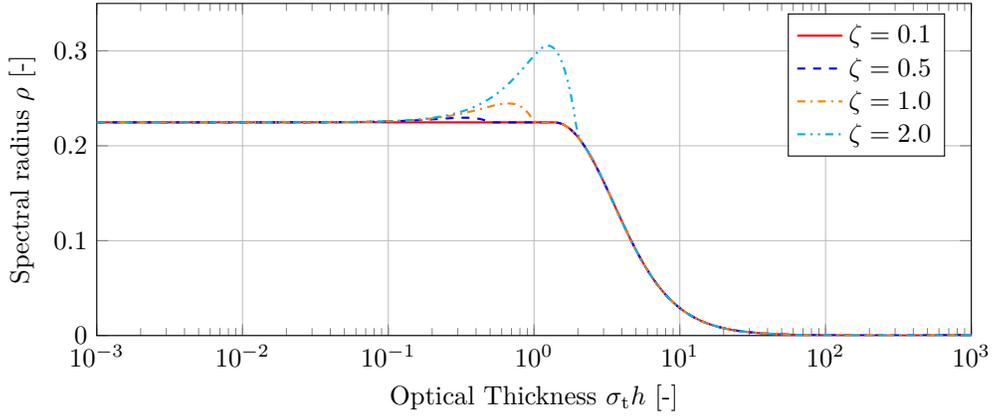
\begin{figure}[tp]
        \begin{minipage}{0.8\textwidth}
            \setlength{\figheight}{6cm}
            \begin{tikzpicture}
	\begin{semilogxaxis}[
    		xmin = 1e-3, xmax = 1e3,
    		ymin = 0, ymax = 0.35,
            ytick = {0,0.1,...,0.4},
    		y tick label style={/pgf/number format/.cd, scaled y ticks = false, fixed},
    		x tick label style={/pgf/number format/.cd, scaled x ticks = false, fixed},,
    		height=\figheight,
    		width=\textwidth,
    		grid=major,
    		legend pos = north east,
    		legend columns = 1,
            cycle list name = color list,
    		xlabel = {Optical Thickness \(\sigt h \) [-]},
    		ylabel = {Spectral radius \(\rho\) [-]}
		]

		\addplot [red, thick, mark phase = 0] table [x = xsh, y = zeta_0_1, col sep=comma] {data/fourier_analysis/infinite_tau_zetas.csv};
        \addlegendentry{\(\zeta = 0.1\)}

		\addplot [blue, thick, dashed] table [x = xsh, y = zeta_0_5, col sep=comma] {data/fourier_analysis/infinite_tau_zetas.csv};
        \addlegendentry{\(\zeta = 0.5\)}

        \addplot [orange, thick, dashdotted] table [x = xsh, y = zeta_1_0, col sep=comma] {data/fourier_analysis/infinite_tau_zetas.csv};
        \addlegendentry{\(\zeta = 1.0\)}

        \addplot [cyan, thick, dashdotdotted] table [x = xsh, y = zeta_2_0, col sep=comma] {data/fourier_analysis/infinite_tau_zetas.csv};
        \addlegendentry{\(\zeta = 2.0\)}

	\end{semilogxaxis}
\end{tikzpicture}
        \end{minipage}
        \caption{Spectral radius for \(c=1\) as function of the optical cell thickness and the threshold parameter \(\zeta\) for the \(\tau\) formulation in an infinite homogeneous material.}
        \label{fig:fourier_analysis_saaft}
    \end{figure}

    The spectral radius of the \(\tau\)-formulations as a function of the parameter \(\zeta\) is shown in \cref{fig:fourier_analysis_saaft} for the infinite, homogeneous case. The spectral radius is dependent on \(\zeta\), and for \(\zeta > 0.1\) the results showed an increase spectral radius around a cell width of one mean-free path~\cite{wang_diffusion_2014}.
    
    \subsection{Diffusion coefficient in voids}
    
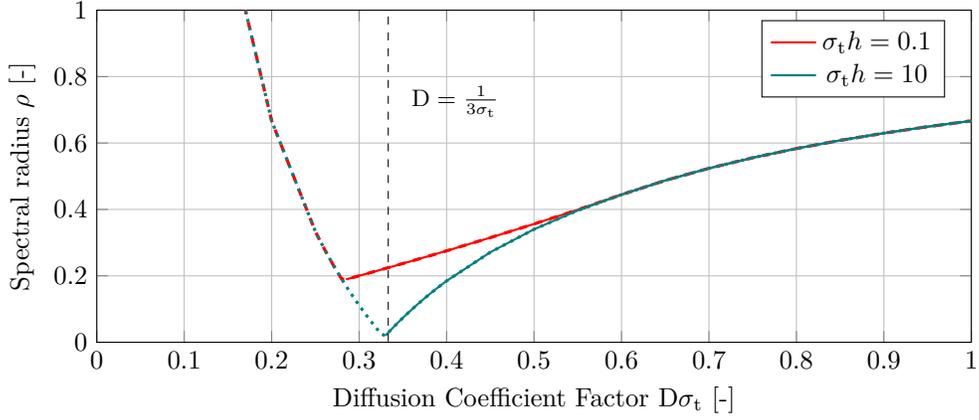
\begin{figure}[p]
        \begin{minipage}{0.8\textwidth}
            \setlength{\figheight}{6cm}
            \begin{tikzpicture}
	\begin{axis}[
		xmin = 0, xmax = 1,
		ymin = 0.0, ymax = 1.0,
        ytick = {-1.0,-0.8,...,1.0},
		y tick label style={/pgf/number format/.cd, scaled y ticks = false, fixed},
		x tick label style={/pgf/number format/.cd, scaled x ticks = false, fixed},,
		height=\figheight,
		width=\textwidth,
		grid=major,
		legend pos = north east,
		legend columns = 1,
		xlabel = {Diffusion Coefficient Factor \(\DC\sigt\) [-]},
		ylabel = {Spectral radius \(\rho\) [-]}
		]
		
        \addplot [red, thick] table [x = D, y = mfp_0.1, col sep=comma, restrict y to domain=0:1.05] {data/fourier_analysis/constd_single_cell_pure_scatterer_results.csv};
            \addplot [red, dashed, very thick, forget plot] table [x = D, y expr = {abs(\thisrow{mfp_0.1})}, col sep=comma] {data/fourier_analysis/constd_single_cell_pure_scatterer_results.csv};
        \addlegendentry{\(\sigt h = 0.1\)}
        
        \addplot [teal, thick] table [x = D, y = mfp_10.0, col sep=comma, restrict y to domain=0:1.05] {data/fourier_analysis/constd_single_cell_pure_scatterer_results.csv};
            \addplot [teal, dotted, very thick, forget plot] table [x = D, y expr = {abs(\thisrow{mfp_10.0})}, col sep=comma] {data/fourier_analysis/constd_single_cell_pure_scatterer_results.csv};
        \addlegendentry{\(\sigt h = 10\)}     
           
        \addplot [black, dashed, below, forget plot] coordinates {(0.3333, -1) (0.3333, 1.5)};
        
        \pgfplotsset{
            after end axis/.code={
                \node[black,below] at (axis cs:0.41,0.8){\small{\(\DC = \frac{1}{3\sigt}\)}};
            }
        }

	\end{axis}
\end{tikzpicture}
        \end{minipage}
        \caption{Spectral radius for \(c=1\) as function of the diffusion coefficient factor in an infinite homogeneous medium for two optical thicknesses (a dotted line indicates a negative eigenvalue).}
        \label{fig:fourier_analysis_diffusion2}
    \end{figure}

    The classical formulation of the diffusion coefficient (\cref{eq:diffusion_coefficient}) is unbounded in voids. If we consider \cref{eq:drift_diffusion}, we see that in the case of spatial and iterative convergence the diffusion terms cancel, however the diffusion coefficient has a strong influence on the spectral radius as shown in \cref{fig:fourier_analysis_diffusion2}. In this figure we varied the factor \(1/3\) of the diffusion coefficient for an optically thin and an optically thick case, and calculated the corresponding spectral radius. In this plot, dotted lines indicate the absolute value of a negative eigenvalues, which means the error (\cref{eq:si_error}) oscillates around the exact solution. The result showed that diffusion coefficients larger than the local diffusion coefficient slowly increase the spectral radius. For smaller diffusion coefficients the increase of the spectral radius is rapid with small variations from the optimal value. Therefore we need a method to obtain a diffusion coefficient for voids with only slight effects on the diffusion coefficient for material regions.
    
    We chose to use a non-local definition of the diffusion coefficient, which is close to the local diffusion coefficient in optical thick cells and well limited in optical thin cells. The derivation was first proposed by Morel \cite{morel_non-local_2007,morel_alternative_2010} and later studied by Larsen and Trahan \cite{larsen_2-d_2009,trahan_3-d_2011} and Schunert~\cite{schunert_using_2016}. Larsen and Morel~\cite{larsen_nonlocal_2017} extended the theory recently to anisotropic scattering. \par
    
    The non-local diffusion coefficient is a \(3\times3\) tensor
    \begin{equation}
        \DCNL \equiv \frac{1}{4\pi}\intsp\direction\int_{0}^{\ell\left(x, -\direction\right)} \e{-\int_{0}^{s} \sigt\left(x - s'\direction\right)\dx[s']} \direction \domg
    \end{equation}
    where \(\ell\) is the distance to the boundary in direction \direction. With the line integral operator \(\left(\direction \cdot \del + \sigt \right)^{-1}\), which can be found for an arbitrary function \(h\) using the method of characteristics as
    \begin{equation} \label{eq:nldc_integral}
        \left(\direction \cdot \del + \sigt \right)^{-1} h\left(x, \direction \right)
        = \int_{0}^{\ell\left(x, -\direction\right)} \e{-\int_{0}^{s} \sigt\left(x - s'\direction\right)\dx[s']} g\left(x - s\direction\right)\dx[s]
    \end{equation}
    this equation can be expressed as
    \begin{equation} \label{eq:def_lonlocal_dii}
        D_{ij} \equiv \frac{1}{4\pi} \intsp \left(\direction\cdot\vec{e}_i\right)\left(\direction\cdot\vec{e}_j\right)g\left(\direction\right)\domg.
    \end{equation}
    In this equation \(g\left(\direction\right)\) is the solution to an auxiliary transport problem
    \begin{subequations} \label{eq:nldc_aux_system}
        \begin{equation}
            \direction\cdot\del g + \sigt g = 1
        \end{equation}
        with the vacuum and reflective boundary conditions (\(\direction_\mathrm{R}\) is the reflected angle for \direction)
        \begin{align}
            g\left(x_b, \direction \right) &= 0, & \forall x_b \in \boundary_\mathrm{V}, \quad \direction \cdot \normal < 0 \\
            g\left(x_b, \direction \right) &= g\left(x_b, \direction_\mathrm{R} \right), & \forall x_b \in \boundary_\mathrm{R}, \quad \direction \cdot \normal < 0.
        \end{align}
    \end{subequations}
    This equation can be easily solved using any technique to solve a transport equation. In this study we obtain the non-local diffusion tensor from a WLS solve. Morel proposed originally reflective boundary conditions for the whole problem, however in this paper the actual boundary conditions of the problem were used. Note that the equation does not have a scattering source, therefore no source iterations are necessary. The result is well defined in finite voids. \par

    For an infinite homogeneous medium, the non-local diffusion coefficient reduced to the classical local diffusion coefficient. This can easily be shown by using the equilibrium solution
    \begin{equation}
        g = \frac{1}{\sigt}
    \end{equation}
    in \cref{eq:def_lonlocal_dii}. The result is a matrix with the local diffusion coefficient on the main diagonal and all other entries zero. \par
    
    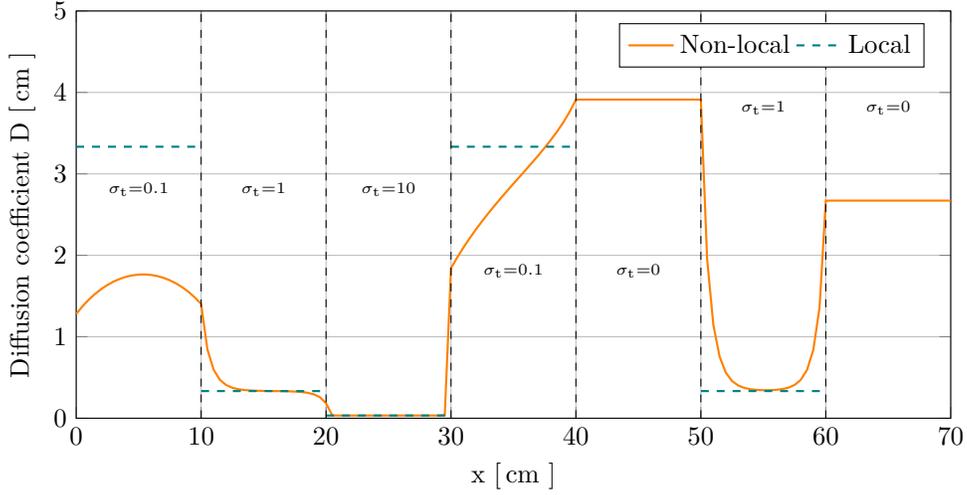
\begin{figure}[tp]
        \begin{minipage}{0.8\textwidth}
            \setlength{\figheight}{7cm}
            \begin{tikzpicture}
    \begin{axis}[
            xmin = 0.0, xmax = 70,
            ymin = 0.0, ymax = 5.0,
            ytick = {0.0,1.0,...,5.0},
            height=\figheight,
            width=\textwidth,
            grid=major,
            legend pos = north east,
            legend columns = 2,
            xlabel = {x [\cm]},
            ylabel = {Diffusion coefficient \DC[] [\cm]},
            mark options={solid, scale = 0.75}
        ]
        
        \addplot [orange, thick] table [x = x, y = D, col sep=comma] {data/fourier_analysis/nl_coefficient.csv};
        \addlegendentry{Non-local}
        
        \addplot [teal, dashed, thick] coordinates {( 0.0, 3.3333) (10, 3.3333)};
        \addplot [teal, dashed, thick] coordinates {(10.0, 0.3333) (20, 0.3333)};
        \addplot [teal, dashed, thick] coordinates {(20.0, 0.0333) (30, 0.0333)};
        \addplot [teal, dashed, thick] coordinates {(30.0, 3.3333) (40, 3.3333)};
        \addplot [teal, dashed, thick] coordinates {(50.0, 0.3333) (60, 0.3333)};
        \addlegendentry{Local}
        
        \addplot [black, below, dashed] coordinates {(10, 0.0) (10, 15.0)};
        \addplot [black, below, dashed] coordinates {(20, 0.0) (20, 15.0)};
        \addplot [black, below, dashed] coordinates {(30, 0.0) (30, 15.0)};
        \addplot [black, below, dashed] coordinates {(40, 0.0) (40, 15.0)};
        \addplot [black, below, dashed] coordinates {(50, 0.0) (50, 15.0)};
        \addplot [black, below, dashed] coordinates {(60, 0.0) (60, 15.0)};
        
        \pgfplotsset{
            after end axis/.code={
                \node[black,below] at (axis cs:5, 3){\small{\(\substack{\sigt[]= 0.1}\)}};
                \node[black,below] at (axis cs:15, 3){\small{\(\substack{\sigt[]= 1}\)}};
                \node[black,below] at (axis cs:25, 3){\small{\(\substack{\sigt[]= 10}\)}};
                \node[black,below] at (axis cs:35, 2){\small{\(\substack{\sigt[]= 0.1}\)}};
                \node[black,below] at (axis cs:45, 2){\small{\(\substack{\sigt[]= 0}\)}};
                \node[black,below] at (axis cs:55, 4){\small{\(\substack{\sigt[]= 1}\)}};
                \node[black,below] at (axis cs:65, 4){\small{\(\substack{\sigt[]= 0}\)}};
            }
        }
    
    \end{axis}
\end{tikzpicture}
        \end{minipage}
        \caption{Non-local diffusion coefficient for a problem with several material regions (cross sections in \Xsunit) in comparison to the local diffusion coefficient.}
        \label{fig:fourier_analysis_nonlinear_coefficient}
    \end{figure}
    
    An example for the non-local diffusion coefficient is shown in \cref{fig:fourier_analysis_nonlinear_coefficient}. The non-local diffusion coefficient is well limited in voids, but the actual value is dependent on the adjacent material regions. If the regions next to a void are optical thin, the value is larger than next to optical thick regions. For optical thick regions the non-local diffusion coefficient settles fast to the value of the local coefficient. In optical thin regions, this equilibrium is not reached. \par
    
    It is of interest how the non-local diffusion coefficient compares in voids to other possible coefficient. One alternative is to limit the use of the non-local diffusion coefficients to optically thin regions. The resulting diffusion coefficient is referred to as the joined coefficient and defined as
    \begin{align} \label{eq:def_diffusion_joined}
    \DC \equiv \begin{cases}
    \DC[\mathrm{local}] & \sigt h \ge \zeta_{\DC} \\
    \DC[\mathrm{nonlocal}] & \sigt h < \zeta_{\DC}
    \end{cases}.
    \end{align}
    We chose \(\zeta_{\DC} < \tento{-3}\) for our test cases, since for this setting the void region always uses the non-local and the material region always uses the local diffusion coefficient. All larger \(\zeta_{\DC}\) would give a combination of these results and the pure non-local case (\(\zeta_{\DC} = \infty\)). The third option is to limit the local diffusion coefficient \cref{eq:diffusion_coefficient} by
    \begin{equation} \label{eq:def_local_diff_limited}
    \DC = \min\left(\frac{1}{3\sigt}, \DC[\mathrm{max}]\right)
    \end{equation}
    where \DC[\mathrm{max}] is a constant. We refer to this as the limited diffusion coefficient. We considered several cases for \DC[\mathrm{max}] up to 1000. \par

\subsection{Analysis of heterogeneous test problems for NDA schemes}
    \begin{table}[p]
        \caption{Void test cases for the periodic two region Fourier analysis, each region uses 2 cells.}
        \label{tab:fa_input_cases_void}
        \renewcommand{\arraystretch}{1.5}
        \begin{tabular}{cc@{\hskip 1cm}ll@{\hskip 1cm}ll}
            \toprule
            \multicolumn{2}{c@{\hskip 1cm}}{Case} &   \multicolumn{2}{c@{\hskip 1cm}}{Region 1}    &        \multicolumn{2}{c}{Region 2}         \\ \midrule
            &                a                & \(\sigt[1] = 1.0\Xsunit \) & \(c_1 = 1.0 \)    & \(\sigt[2] = 0.0\Xsunit \) & \(c_2 = 0.0 \) \\
            &                b                & \(\sigt[1] = 1.0\Xsunit \) & \(c_1 = 0.9999 \) & \(\sigt[2] = 0.0\Xsunit \) & \(c_2 = 0.0 \) \\
            &                c                & \(\sigt[1] = 1.0\Xsunit \) & \(c_1 = 0.99 \)   & \(\sigt[2] = 0.0\Xsunit \) & \(c_2 = 0.0 \) \\
            &                d                & \(\sigt[1] = 1.0\Xsunit \) & \(c_1 = 0.9 \)    & \(\sigt[2] = 0.0\Xsunit \) & \(c_2 = 0.0 \) \\ \bottomrule
        \end{tabular}
        \renewcommand{\arraystretch}{1}
    \end{table}

    The combination of the non-local diffusion coefficient and either one of the three drift-vector formulations Current, Combined or \(\tau\) allows us now to use the NDA in regions with voids. The heterogeneous Fourier analysis is now used to study the effects of void regions on the convergence and stability of the NDA algorithm with these modifications. For this purpose we used four test cases shown in \cref{tab:fa_input_cases_void}. Each case consists of four cells with cell width \(h\). The geometry is periodic, but we don't assume a periodic solution. For all test cases the cell size \(h\) was varied in a range from \tento{-3}cm to \tento{3}cm. For every \(h\), the eigenvalue for 200 frequencies between \(\lambda = \tento{-4}\) and \(\lambda = \pi\) were calculated. To resolve the eigenvalues for small \(h\) better, more points were calculated for smaller \(\lambda\), where the spectral radius peaks with a sharp peak (see \cref{fig:fourier_analysis_frequency} in \cref{sec:fourier_numerical}). To prevent numerical instabilities, the lower cutoff for \(\lambda\) was raised for larger cell thicknesses \(h\), since the peak shifts towards higher frequencies. The non-local diffusion coefficient was calculated using a WLS solver on the same mesh. \par

    \begin{table}
        \caption{Eigenvalues with the largest magnitude for the void test cases for the NDA formulations (\cref{tab:fa_input_cases_void}) using the non-local diffusion coefficient WLS transport solves.}
        \label{tab:fourier_analysis_nda_void}
        \begin{tabular}{cc@{\hskip 20pt}l@{\hskip 20pt}r@{\hskip 20pt}r@{\hskip 20pt}r}
        	\toprule
        	        Case         & \multicolumn{1}{c}{\(c\)} &           \multicolumn{1}{c}{Current}            &           \multicolumn{1}{c}{Combined}           &                  \multicolumn{1}{c}{\(\tau\)}                  \\ \midrule
        	      a &  1.0000    &  0.9947 &     0.9939 &      -9910.0000 \\
      b &  0.9999    &  0.9357 &     0.6468 &     -313.1000 \\
      c &  0.99      &  0.8540 &     0.4625 &     -11.3900 \\
      d &  0.9       &  0.6159 &     0.3658 &     -1.2360 \\\bottomrule
 &
        \end{tabular}
    \end{table}

     The results for all void cases (\cref{tab:fa_input_cases_void}) using the three schemes are shown in \cref{fig:fourier_nda_void}. For all \(c\) the Combined formulation (\cref{fig:fourier_nda_void_combined}) showed a smaller spectral radius than the Current formulation (\cref{fig:fourier_nda_void_current}). Nevertheless also the spectral radius of the Combined formulation goes to one for large \(h\) and \(c=1\). However, reducing the scattering ratio only to \(c = 0.9999\) reduced the maximum spectral radius significantly as shown in \cref{tab:fourier_analysis_nda_void}. The \(\tau\) scheme using a WLS transport solution was not unconditionally stable as shown in \cref{fig:fourier_nda_void_saaft}. Based on these results, the combined current formulation is the best choice for the NDA WLS in voids. It provides unconditionally stable and efficient acceleration for all physical relevant problems. The small oscillations for small \(h\) are results of the algorithm not finding the highest eigenvalue due to small peaks. \par

    \begin{table}
        \caption{Comparison of the eigenvalues with the largest magnitude for the void test cases with the combined NDA WLS formulation (\cref{tab:fa_input_cases_void}) using the non-local, joined and limited diffusion coefficients with different parameters.}
        \label{tab:fourier_analysis_nda_void_coefficient}
        \begin{tabular}{c@{\hskip 20pt}l@{\hskip 20pt}rr@{\hskip 20pt}rrrr}
            \toprule
            \multicolumn{1}{c@{\hskip 20pt}}{Case}         & \multicolumn{1}{c@{\hskip 20pt}}{\(c\)} &           \multicolumn{2}{c@{\hskip 20pt}}{Non-Local}            &           \multicolumn{4}{c}{Limited local}  \\
            &                                         & \multicolumn{1}{c}{\(\zeta_{\DC} = \infty\)} & \tento{-3} & \multicolumn{1}{c}{\(\DC[\mathrm{max}] = 1\cm\)} & 10\cm & 100\cm & 1000\cm \\ \midrule
               a &    1.0000 &    0.9939 &    0.9939 &    1.0270 &    0.9988 &    0.9990 &    0.9990 \\
   b &    0.9999 &    0.6468 &    0.6660 &    0.9521 &    0.8947 &    0.9644 &    0.9885 \\
   c &      0.99 &    0.4625 &    0.5354 &    0.6572 &    0.4491 &    0.7223 &    0.8870 \\
   d &       0.9 &    0.3658 &    0.3267 &    0.3301 &    0.1944 &    0.4092 &    0.6564 \\\bottomrule

        \end{tabular}
    \end{table}

    \Cref{tab:fourier_analysis_nda_void_coefficient} shows the spectral radii for the different diffusion coefficients, \cref{eq:def_diffusion_joined,eq:def_local_diff_limited}, in the void test cases (\cref{tab:fa_input_cases_void}). For pure scatterers, all coefficients resulted in a spectral radius of close to 1. The higher limited coefficients had a lower spectral radius in the intermediate \(h\) range, but increased toward \(\rho = 1\) earlier than the non-local coefficient. For \(c = 0.9999\) the non-local coefficient was significantly better than the limited schemes. However for \(c \le 0.9\) the other schemes started to perform better or equal than the pure non-local coefficient as can be seen. But in these cases the spectral radii for the non-local cases were well below 0.4, which results in an efficient acceleration anyways. \par
    
        \begin{figure}[H]
        \begin{minipage}{0.8\textwidth}
            \setlength{\figheight}{7.5cm}
            \begin{tikzpicture}
	\begin{semilogxaxis}[
    		xmin = 1e-3, xmax = 1e3,
    		ymin = 0.0, ymax = 1.0,
            ytick = {-1.0,-0.8,...,1.0},
    		y tick label style={/pgf/number format/.cd, scaled y ticks = false, fixed},
    		x tick label style={/pgf/number format/.cd, scaled x ticks = false, fixed},,
    		height=\figheight,
    		width=\textwidth,
    		grid=major,
    		legend pos = north west,
    		legend columns = 1,
            cycle list name = color list,
    		xlabel = {Cell Thickness \(h\) [\cm]},
    		ylabel = {Spectral radius \(\rho\) [-]}
		]

        \addplot +[red, thick] table [x = x, y expr = {max(\thisrow{nda_void_non_local_current}, -1.1)}, col sep=comma] {data/fourier_analysis/nda_partially_void_c_1_0.csv};
        \addlegendentry{\(c = 1.0\)}

        \addplot +[blue, thick] table [x = x, y expr = {max(\thisrow{nda_void_non_local_current}, -1.1)}, col sep=comma] {data/fourier_analysis/nda_partially_void_c_0_9999.csv};
        \addlegendentry{\(c = 0.9999\)}

        \addplot +[teal, thick] table [x = x, y expr = {max(\thisrow{nda_void_non_local_current}, -1.1)}, col sep=comma] {data/fourier_analysis/nda_partially_void_c_0_99.csv};
        \addlegendentry{\(c = 0.99\)}

        \addplot +[orange, thick] table [x = x, y expr = {max(\thisrow{nda_void_non_local_current}, -1.1)}, col sep=comma] {data/fourier_analysis/nda_partially_void_c_0_9.csv};
        \addlegendentry{\(c = 0.9\)}

	\end{semilogxaxis}
\end{tikzpicture}
            \subcaption{Current drift vector}
            \label{fig:fourier_nda_void_current}
        \end{minipage}
        \begin{minipage}{0.8\textwidth}
            \setlength{\figheight}{7.5cm}
            \begin{tikzpicture}
	\begin{semilogxaxis}[
		xmin = 1e-3, xmax = 1e3,
		ymin = 0.0, ymax = 1.0,
        ytick = {-1.0,-0.8,...,1.0},
		y tick label style={/pgf/number format/.cd, scaled y ticks = false, fixed},
		x tick label style={/pgf/number format/.cd, scaled x ticks = false, fixed},,
		height=\figheight,
		width=\textwidth,
		grid=major,
		legend pos = north west,
		legend columns = 1,
		xlabel = {Cell Thickness \(h\) [\cm]},
		ylabel = {Spectral radius \(\rho\) [-]}
		]
		
        \addplot [red, thick] table [x = x, y expr = {abs(\thisrow{nda_void_non_local_combined})}, col sep=comma] {data/fourier_analysis/nda_partially_void_c_1_0.csv};
        \addlegendentry{\(c = 1.0\)}
        
        \addplot [blue, thick] table [x = x, y expr = {abs(\thisrow{nda_void_non_local_combined})}, col sep=comma] {data/fourier_analysis/nda_partially_void_c_0_9999.csv};
        \addlegendentry{\(c = 0.9999\)}
        
        \addplot [teal, thick] table [x = x, y = nda_void_non_local_combined, col sep=comma] {data/fourier_analysis/nda_partially_void_c_0_99.csv};
        \addlegendentry{\(c = 0.99\)}
        
        \addplot [orange, thick] table [x = x, y = nda_void_non_local_combined, col sep=comma] {data/fourier_analysis/nda_partially_void_c_0_9.csv};
        \addlegendentry{\(c = 0.9\)}
        
	\end{semilogxaxis}
\end{tikzpicture}
            \subcaption{Combined drift vector}
            \label{fig:fourier_nda_void_combined}
        \end{minipage}
        \caption{Spectral radii for the void test cases (\cref{tab:fa_input_cases_void}) using the non-local diffusion coefficient for the NDA schemes and a WLS high order solution.}
    \end{figure}
    
    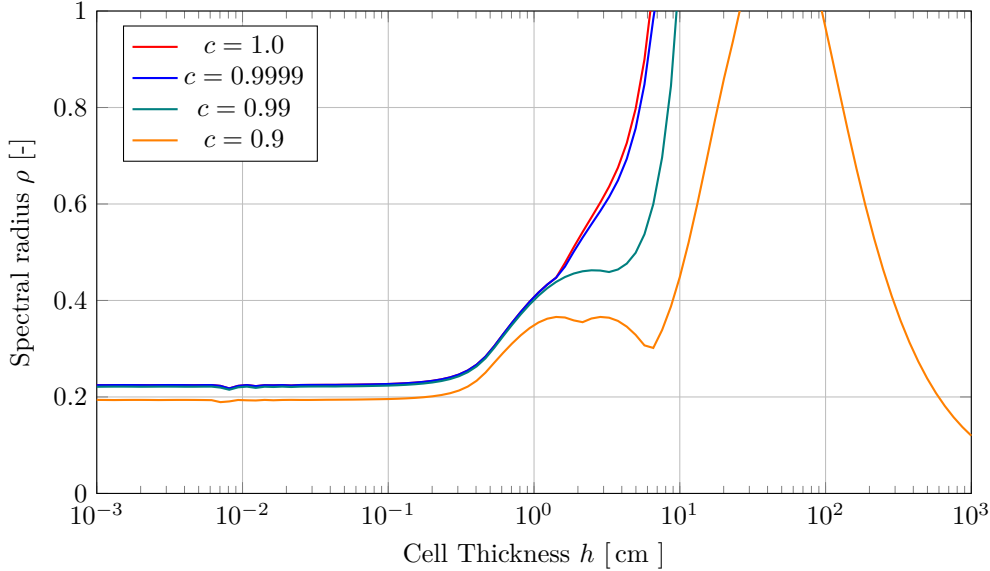
\begin{figure}[H]
        \ContinuedFloat
        \begin{minipage}{0.8\textwidth}
            \setlength{\figheight}{8cm}
            \begin{tikzpicture}
	\begin{semilogxaxis}[
    		xmin = 1e-3, xmax = 1e3,
    		ymin = 0.0, ymax = 1.0,
            ytick = {-1.0,-0.8,...,1.0},
    		y tick label style={/pgf/number format/.cd, scaled y ticks = false, fixed},
    		x tick label style={/pgf/number format/.cd, scaled x ticks = false, fixed},,
    		height=\figheight,
    		width=\textwidth,
    		grid=major,
    		legend pos = north west,
    		legend columns = 1,
            cycle list name = color list,
    		xlabel = {Cell Thickness \(h\) [\cm]},
    		ylabel = {Spectral radius \(\rho\) [-]}
		]

        \addplot +[red, thick] table [x = x, y expr = {abs(max(\thisrow{nda_void_non_local_saaft}, -1.1))}, col sep=comma] {data/fourier_analysis/nda_partially_void_c_1_0.csv};
        \addlegendentry{\(c = 1.0\)}

        \addplot +[blue, thick] table [x = x, y expr = {abs(max(\thisrow{nda_void_non_local_saaft}, -1.1))}, col sep=comma] {data/fourier_analysis/nda_partially_void_c_0_9999.csv};
        \addlegendentry{\(c = 0.9999\)}

        \addplot +[teal, thick] table [x = x, y expr = {abs(max(\thisrow{nda_void_non_local_saaft}, -1.1))}, col sep=comma] {data/fourier_analysis/nda_partially_void_c_0_99.csv};
        \addlegendentry{\(c = 0.99\)}

        \addplot +[orange, thick] table [x = x, y expr = {abs(max(\thisrow{nda_void_non_local_saaft}, -1.1))}, col sep=comma] {data/fourier_analysis/nda_partially_void_c_0_9.csv};
        \addlegendentry{\(c = 0.9\)}

	\end{semilogxaxis}
\end{tikzpicture}
            \subcaption{\(\tau\) drift vector (\(\zeta = 0.1\))}
            \label{fig:fourier_nda_void_saaft}
        \end{minipage}
        \caption[]{Continued.}
        \label{fig:fourier_nda_void}
    \end{figure}
    
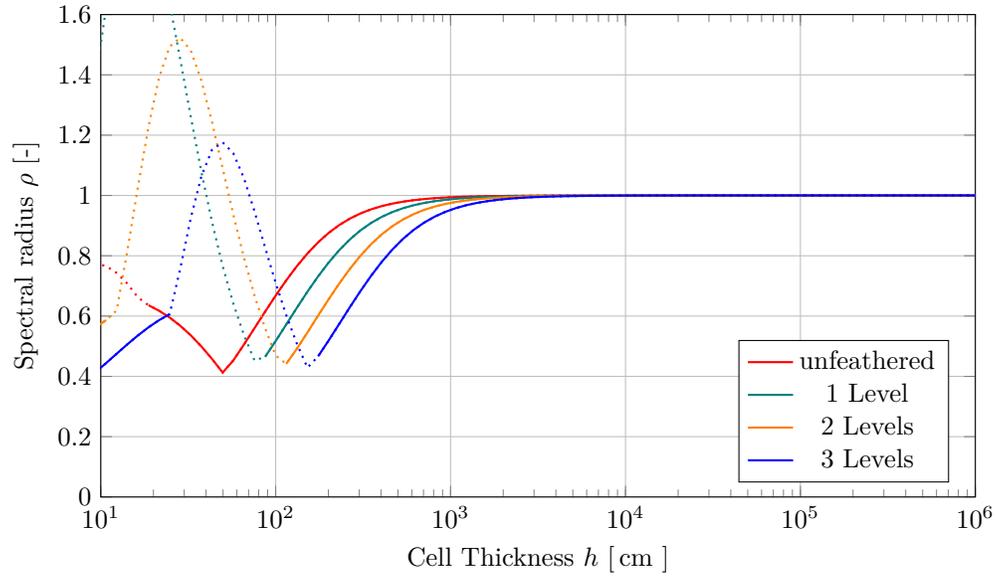
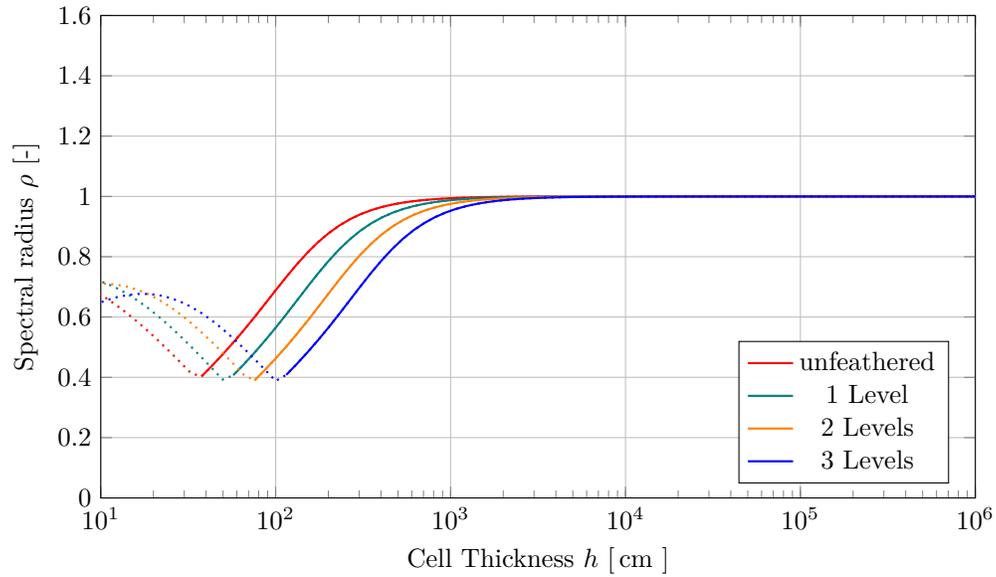
\begin{figure}[p]
        \begin{minipage}{0.8\textwidth}
            \setlength{\figheight}{8cm}
            \begin{tikzpicture}
	\begin{semilogxaxis}[
		xmin = 1e1, xmax = 1e6,
		ymin = 0.0, ymax = 1.6,
        ytick = {-1.0,-0.8,...,1.6},
		y tick label style={/pgf/number format/.cd, scaled y ticks = false, fixed},
		x tick label style={/pgf/number format/.cd, scaled x ticks = false, fixed},,
		height=\figheight,
		width=\textwidth,
		grid=major,
		legend pos = south east,
		legend columns = 1,
        xlabel = {Cell Thickness \(h\) [\cm]},
		ylabel = {Spectral radius \(\rho\) [-]}
		]
        
        \addplot [red, thick] table [x = x, y = feather_0, col sep=comma, restrict y to domain=0:1.05] {data/fourier_analysis/feather_partially_void_c_1_0.csv};
            \addplot [red, dotted, thick, forget plot] table [x = x, y expr = {abs(\thisrow{feather_0})}, col sep=comma] {data/fourier_analysis/feather_partially_void_c_1_0.csv};
        \addlegendentry{unfeathered}
		
        \addplot [teal, thick] table [x = x, y = feather_1, col sep=comma, restrict y to domain=0:1.05] {data/fourier_analysis/feather_partially_void_c_1_0.csv};
            \addplot [teal, dotted, thick, forget plot] table [x = x, y expr = {abs(\thisrow{feather_1})}, col sep=comma] {data/fourier_analysis/feather_partially_void_c_1_0.csv};
        \addlegendentry{1 Level}
        
        \addplot [orange, thick] table [x = x, y = feather_2, col sep=comma, restrict y to domain=0:1.05] {data/fourier_analysis/feather_partially_void_c_1_0.csv};
        \addplot [orange, dotted, thick, forget plot] table [x = x, y expr = {abs(\thisrow{feather_2})}, col sep=comma] {data/fourier_analysis/feather_partially_void_c_1_0.csv};
        \addlegendentry{2 Levels}
        
        \addplot [blue, thick] table [x = x, y = feather_3, col sep=comma, restrict y to domain=0:1.05] {data/fourier_analysis/feather_partially_void_c_1_0.csv};
        \addplot [blue, dotted, thick, forget plot] table [x = x, y expr = {abs(\thisrow{feather_3})}, col sep=comma] {data/fourier_analysis/feather_partially_void_c_1_0.csv};
        \addlegendentry{3 Levels}
        
	\end{semilogxaxis}
\end{tikzpicture}
            \subcaption{Non-local diffusion coefficient}
            \label{fig:fourier_analysis_feather_nonlocal}
        \end{minipage}
        \begin{minipage}{0.8\textwidth}
            \setlength{\figheight}{8cm}
            \begin{tikzpicture}
	\begin{semilogxaxis}[
		xmin = 1e1, xmax = 1e6,
		ymin = 0.0, ymax = 1.6,
        ytick = {-1.0,-0.8,...,1.6},
		y tick label style={/pgf/number format/.cd, scaled y ticks = false, fixed},
		x tick label style={/pgf/number format/.cd, scaled x ticks = false, fixed},,
		height=\figheight,
		width=\textwidth,
		grid=major,
		legend pos = south east,
		legend columns = 1,
        xlabel = {Cell Thickness \(h\) [\cm]},
		ylabel = {Spectral radius \(\rho\) [-]}
		]
        
        \addplot [red, thick] table [x = x, y = feather_0, col sep=comma, restrict y to domain=0:1.05] {data/fourier_analysis/feather_partially_void_c_1_0_op.csv};
            \addplot [red, dotted, thick, forget plot] table [x = x, y expr = {abs(\thisrow{feather_0})}, col sep=comma] {data/fourier_analysis/feather_partially_void_c_1_0_op.csv};
        \addlegendentry{unfeathered}
		
        \addplot [teal, thick] table [x = x, y = feather_1, col sep=comma, restrict y to domain=0:1.05] {data/fourier_analysis/feather_partially_void_c_1_0_op.csv};
            \addplot [teal, dotted, thick, forget plot] table [x = x, y expr = {abs(\thisrow{feather_1})}, col sep=comma] {data/fourier_analysis/feather_partially_void_c_1_0_op.csv};
        \addlegendentry{1 Level}
        
        \addplot [orange, thick] table [x = x, y = feather_2, col sep=comma, restrict y to domain=0:1.05] {data/fourier_analysis/feather_partially_void_c_1_0_op.csv};
        \addplot [orange, dotted, thick, forget plot] table [x = x, y expr = {abs(\thisrow{feather_2})}, col sep=comma] {data/fourier_analysis/feather_partially_void_c_1_0_op.csv};
        \addlegendentry{2 Levels}
        
        \addplot [blue, thick] table [x = x, y = feather_3, col sep=comma, restrict y to domain=0:1.05] {data/fourier_analysis/feather_partially_void_c_1_0_op.csv};
        \addplot [blue, dotted, thick, forget plot] table [x = x, y expr = {abs(\thisrow{feather_3})}, col sep=comma] {data/fourier_analysis/feather_partially_void_c_1_0_op.csv};
        \addlegendentry{3 Levels}
        
	\end{semilogxaxis}
\end{tikzpicture}
            \subcaption{Joined diffusion coefficient}
            \label{fig:fourier_analysis_feather_joined}
        \end{minipage}
        \caption{Spectral radius for \(c=1\) as function of the cell thickness for different levels of feathering using the nonlocal and the joined diffusion coefficient (dotted line indicates negative eigenvalues).}
        \label{fig:fourier_analysis_feather}
    \end{figure}

    For optical thick cells and pure scatterers the void-compatible NDA scheme loses effectiveness. This is caused by an interface between an optically very thick cell and a very thin cell. To improve convergence we studied the effect of introducing cells of intermediate optical thickness. The last interface cells in the material on both sides were increasingly refined towards the void, creating a series of cells with decreasing optical thickness towards the void region. Every level of feathering means that the cells next to the void is divided by two, hence a level three feathering produces an interface with cells of \(\sigt h /2\), \(\sigt h/4\) and two with \(\sigt h/8\) thickness. In this study we modified the cross section to maintain a regular mesh.  \Cref{fig:fourier_analysis_feather} shows that this procedure moved the practical optical thickness, for which the scheme lost effectiveness to optically thicker cells. This, however, came with the price of having more spatial cells in the problem. The implemented feathering scheme was only intended for a test and is by no means optimal.
    
    Nevertheless, the test showed that the use of the non-local diffusion coefficient can lead to unstable systems (\cref{fig:fourier_analysis_feather_nonlocal}) if the coefficient was calculated on a mesh that is not sufficiently refined. Using the WLS transport equation resulted in oscillations at the void-material interface in the diffusion coefficient for unresolved thick cells, which became negative for thicknesses between 10 and 100, degrading the convergence rate. This results was expected, since the diffusion coefficient is being inaccurately computed. Using the joined diffusion coefficient (\cref{eq:def_diffusion_joined}), these oscillations can be eliminated and the scheme converges again as can be seen in \cref{fig:fourier_analysis_feather_joined}. This shows the importance to obtain a good approximation of the non-local diffusion coefficient.
    
\subsection{Numerical Fourier Analysis} \label{sec:fourier_numerical}  

    To verify our finding we use a numerical code to obtain the spectral radii and compare it to the values obtained from the Fourier analysis. Since the NDA method is a non-linear method, it is not possible to use the traditional method to converge against a zero solution. Hence we introduce a source \(q = \siga\) to obtain a constant solution \(\phi = 1\) everywhere in the problem. This limits the number of iterations we can perform before having problems with machine accuracy. especially for small spectral radii. The calculations were randomly initialized with values uniformly distributed between 0 and 10 to cover all frequencies. Of interest were the number of iterations after which the error was reduced by a factor of \tento{-6} to the initial random guess and the spectral radius, which was obtained as the ratio of the errors of the last iteration to the previous iteration. We ran 10 samples for every case and took the average over these for the spectral radii and the number of iterations to limit the influence of a specific initial guess. \par
    
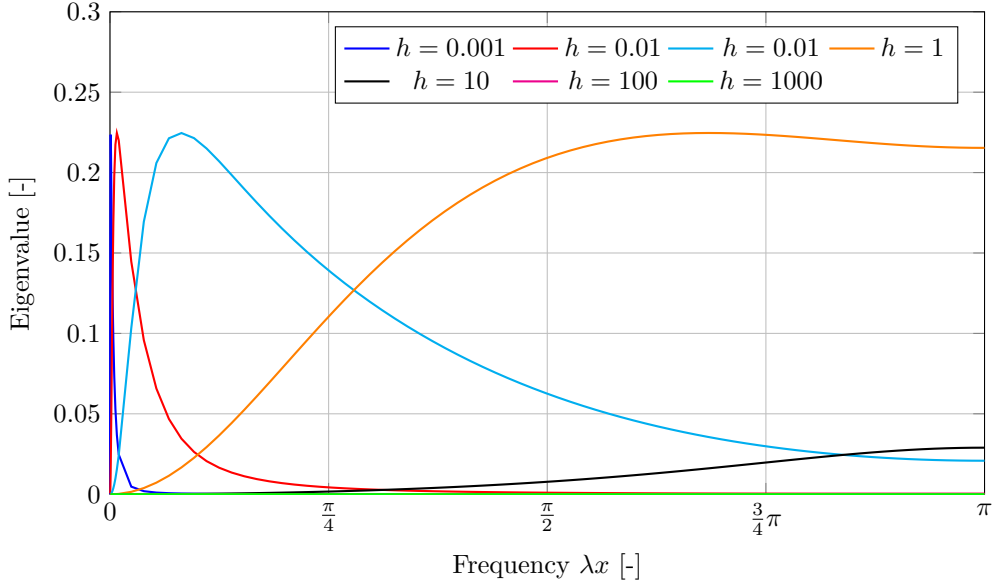
\begin{figure}
        \begin{minipage}{0.8\textwidth}
            \setlength{\figheight}{8cm}
            \begin{tikzpicture}
	\begin{axis}[
    		xmin = 0, xmax = 3.141592654,
    		ymin = 0.0, ymax = 0.3,
            xtick={0, 0.785398163, 1.5708, 2.35619449, 3.14159},
            xticklabels={0, \(\frac{\pi}{4}\), \(\frac{\pi}{2}\), \(\frac{3}{4}\pi\), \(\pi\)},
    		y tick label style={/pgf/number format/.cd, scaled y ticks = false, fixed},
    		height=\figheight,
    		width=\textwidth,
    		grid=major,
    		legend pos = north east,
    		legend columns = 4,
            xlabel = {Frequency \(\lambda x\) [-]},
    		ylabel = {Eigenvalue [-]}
		]
        
        \addplot [blue, thick] table [x = lambda, y = h_0.001, col sep=comma] {data/fourier_analysis/frequencies.csv};
        \addlegendentry{\(h = 0.001\)}
        
        \addplot [red, thick] table [x = lambda, y = h_0.01, col sep=comma] {data/fourier_analysis/frequencies.csv};
        \addlegendentry{\(h = 0.01\)}
        
        \addplot [cyan, thick] table [x = lambda, y = h_0.1, col sep=comma] {data/fourier_analysis/frequencies.csv};
        \addlegendentry{\(h = 0.01\)}
        
        \addplot [orange, thick] table [x = lambda, y = h_1, col sep=comma] {data/fourier_analysis/frequencies.csv};
        \addlegendentry{\(h = 1\)}
        
        \addplot [black, thick] table [x = lambda, y = h_10, col sep=comma] {data/fourier_analysis/frequencies.csv};
        \addlegendentry{\(h = 10\)}
        
        \addplot [magenta, thick] table [x = lambda, y = h_100, col sep=comma] {data/fourier_analysis/frequencies.csv};
        \addlegendentry{\(h = 100\)}
        
        \addplot [green, thick] table [x = lambda, y = h_1000, col sep=comma] {data/fourier_analysis/frequencies.csv};
        \addlegendentry{\(h = 1000\)}
		
%
%
        
	\end{axis}
\end{tikzpicture}
        \end{minipage}
        \caption{Distribution of the eigenvalues as a function of the frequency for all cell thicknesses.}
        \label{fig:fourier_analysis_frequency}
    \end{figure}

    The periodic boundary condition requires that on a mesh with \(N\) cells the scalar flux satisfies the condition
    \begin{equation} \label{eq:nfa_periodic}
        \phi_{0} = \phi_{N + 1}.
    \end{equation}
    This limits the frequencies a mesh with the regular size \(h\) can support, since for all frequencies \(\lambda\) must then hold
    \begin{align}
        \phi_{0} &= \phi_{N + 1} \e{\img \lambda N h} \notag \\
                 &= \phi_{0} \e{\img \lambda N h},
    \end{align}
    which is only true if
    \begin{equation}
        \lambda = \frac{k\pi}{N h} \qquad k \in \mathbb{N}_0.
    \end{equation}
    Therefore a mesh with \(N\) cells has only discrete frequencies, but the analytic Fourier analysis gives the spectral radius over all frequencies. To obtain a better comparison between the analytic and the numerical Fourier analysis, we restricted the frequencies in the analytic Fourier analysis to the frequencies supported by the selected mesh. \Cref{tab:fa_computational_infinite} shows the comparison of the spectral radii from the analytic Fourier analysis with all frequencies (Analytic) to the restricted analytic and the average computational spectral radii for several mean free path on two different meshes. The results show a good agreement between the restricted spectral radii and the observed ones for both meshes. Furthermore the fine mesh with 1000 cells agrees well with the analytic Fourier analysis except for the smallest \(h\). The coarser mesh has larger differences for small cell thicknesses \(h \le 0.01\). The reason that the restricted and numerical spectral radii for small \(h\) is smaller than the predicted is that the eigenvalue peak for these cases is limited to a small range of frequencies as shown in \cref{fig:fourier_analysis_frequency}. The discrete frequencies cannot resolve these small peaks. \par

    \begin{table}
    \caption{Average computational spectral radii using the Eddington formulation for the infinite Fourier analysis with \(c = 0.9999\) compared to the analytical Fourier analysis with all frequencies (Analytic) and with frequencies restricted to the supported ones on the corresponding mesh (Restricted).}
    \label{tab:fa_computational_infinite}
    \begin{tabular}{rr@{\hskip 25pt}rr@{\hskip 25pt}rr}
    	\toprule
    	\multicolumn{1}{c@{\hskip 18pt}}{Cell} & \multirow{2}{*}{Analytic} & \multicolumn{2}{c@{\hskip 25pt}}{100 Mesh Cells} & \multicolumn{2}{c@{\hskip 25pt}}{1000 Mesh Cells} \\
    	         \multicolumn{1}{c}{Thickness} &                           & Restricted &                           Numerical & Restricted &                            Numerical \\ \midrule
    	                                 0.001 &                    0.2236 &     0.0066 &                              0.0063 &     0.1649 &                               0.1634 \\
    	                                  0.01 &                    0.2246 &     0.1631 &                              0.1566 &     0.2246 &                               0.2144 \\
    	                                   0.1 &                    0.2246 &     0.2246 &                              0.2124 &     0.2246 &                               0.2135 \\
    	                                     1 &                    0.2246 &     0.2246 &                              0.2189 &     0.2246 &                               0.2193 \\
    	                                    10 &                    0.0289 &     0.0289 &                              0.0255 &     0.0289 &                               0.0260 \\
    	                                   100 &                    0.0003 &     0.0003 &                              0.0002 &     0.0003 &                               0.0002 \\
    	                                  1000 &                  < 0.0001 &   < 0.0001 &                            < 0.0001 &   < 0.0001 &                             < 0.0001 \\ \bottomrule
    \end{tabular}
    \end{table}
    
    \begin{table}
        \caption{Average computational spectral radii and average number of NDA iterations for the infinite Fourier analysis with \(c = 0.9999\) compared to the analytical Fourier analysis for the Eddington and Current Formulation on a mesh with 1000 Cells.}
        \label{tab:fa_computational_infinite_current}
        \begin{tabular}{r@{\hskip 15pt}rrr@{\hskip 18pt}rrr}
        	\toprule
        	     \multicolumn{1}{c@{\hskip 15pt}}{Cell} & \multicolumn{3}{c@{\hskip 18pt}}{Eddington} & \multicolumn{3}{c@{\hskip 18pt}}{Current} \\
        	\multicolumn{1}{c@{\hskip 18pt}}{Thickness} & Analytic & Numerical &           Iterations & Analytic & Numerical &         Iterations \\ \midrule
        	                                      0.001 &   0.2236 &    0.1634 &                  4.6 &   0.2236 &    0.1630 &                4.5 \\
        	                                       0.01 &   0.2246 &    0.2144 &                  6.0 &   0.2246 &    0.2123 &                5.9 \\
        	                                        0.1 &   0.2246 &    0.2135 &                  6.3 &   0.2259 &    0.2143 &                6.2 \\
        	                                          1 &   0.2246 &    0.2193 &                  7.0 &   0.3723 &    0.3651 &               10.0 \\
        	                                         10 &   0.0289 &    0.0260 &                  3.0 &   0.9602 &    0.9590 &              202.6 \\
        	                                        100 &   0.0003 &    0.0002 &                  2.0 &   0.7996 &    0.7943 &               39.1 \\
        	                                       1000 & < 0.0001 &  < 0.0001 &                  1.0 &   0.0385 &    0.0346 &                3.0 \\ \bottomrule
        \end{tabular}
    \end{table}
    
    The comparison between the spectral radii and average number of iterations for the Eddington and Current formulation, using the fine mesh with 1000 cells, can be seen in \cref{tab:fa_computational_infinite_current}. The numerical analysis clearly showed the predicted increase of the spectral radius for optical thick cells for the Current formulation. Note that for \(c=0.9999\) the spectral radius of the Current formulation does not go asymptotically towards one as it does for \(c=1\), but peaks at approx. \(h=10\) and goes to zero for optical very thick cells. The results demonstrates that this increase in the spectral radius causes the a number of required iterations to be two magnitudes larger than for the Eddington formulation. \par
    
    \begin{table}
        \caption{Average computational spectral radii and average number of NDA iterations for the two region void problem with \(c=0.9999\) (Case 4b, \cref{tab:fa_input_cases_void}) for Combined Formulation using the non-local and the joined diffusion coefficient compared to the corresponding analytical Fourier analysis.}
        \label{tab:fa_computational_void}
        \begin{tabular}{r@{\hskip 15pt}rrr@{\hskip 18pt}rrr}
        	\toprule
        	     \multicolumn{1}{c@{\hskip 18pt}}{Cell} & \multicolumn{3}{c@{\hskip 18pt}}{Non-local Coefficient} & \multicolumn{3}{c@{\hskip 18pt}}{Joined Coefficient} \\
        	\multicolumn{1}{c@{\hskip 15pt}}{Thickness} & Analytic & Numerical &                       Iterations & Analytic & Numerical &                    Iterations \\ \midrule
        	                                      0.001 &   0.2233 &    0.1769 &                              4.3 &   0.4777 &    0.1258 &                           3.7 \\
        	                                       0.01 &   0.2245 &    0.2119 &                              5.8 &   0.4781 &    0.3277 &                           6.6 \\
        	                                        0.1 &   0.2268 &    0.2142 &                              6.0 &   0.4799 &    0.4671 &                          10.8 \\
        	                                          1 &   0.4070 &    0.4267 &                             11.3 &   0.5751 &    0.5661 &                          18.0 \\
        	                                         10 &   0.4785 &    0.4531 &                             14.1 &   0.5452 &    0.5379 &                          16.6 \\
        	                                        100 &   0.6484 &    0.7867 &                             29.1 &   0.6377 &    0.6297 &                          21.0 \\
        	                                       1000 &   0.0413 &    0.0339 &                              4.0 &   0.0413 &    0.0340 &                           4.0 \\ \bottomrule
        \end{tabular}
    \end{table}
    
     We performed the same calculations for test case b (\cref{tab:fa_input_cases_void}), a void case with a scattering region with \(c = 0.9999\) using the Combined Formulation with the non-local diffusion coefficient. We used 2000 cells for the periodic problem, 2 cells per region as we did for the analytic Fourier analysis and repeated this 500 times to obtain a decent sized mesh. Again we ran 10 samples and took the average. The results as shown in \cref{tab:fa_computational_void} showed good agreement to the predicted spectral radii except for \(h=100\). At this thickness the numerical spectral radius was significantly higher than the analytical, which cannot be explained by the limitation of frequencies. However the cells are too thick for a reliable calculations of the non-local diffusion coefficient, which gave negative results in the material region, causing a degrading of convergence. To prove this, the diffusion coefficient was switched to the joined diffusion coefficient (\cref{eq:def_diffusion_joined}) with \(\zeta_D = \tento{-3}\). This used the local diffusion coefficient in the material region and hence avoids the increase in the spectral radius as can be seen in the right half of \cref{tab:fa_computational_void}. Since the non-local diffusion coefficient is constant in the void region, the result in the void can be used without problems. \par
     
\section{Numerical Results} \label{ch:results}

\subsection{Reed's Problem}

    \setlength{\figheight}{5cm}
    \begin{figure}[th]
        \begin{minipage}{0.8\textwidth}
            \setlength{\figheight}{8cm}
            \begin{tikzpicture}
	\begin{axis}[
            xmin = 0, xmax = 8,
            ymin = 0, ymax = 3,
            y tick label style={/pgf/number format/.cd, scaled y ticks = false, fixed},
            x tick label style={/pgf/number format/.cd, scaled x ticks = false, fixed},,
            xtick = {0, 1, ..., 8},
            ytick = {0, 1, 2, ..., 6},
            height=\figheight,
            width=1.0\textwidth,
            grid=major,
              	legend pos = north west,
            legend columns = 1,
            xlabel = {\(x\) [\cm] (64 cells)},
            ylabel = {Scalar flux \(\phi\) [\sfluxunit]},
            mark options={solid, scale = 0.4}
		]

		\addplot [red, thick] table [x = x, y = reference, col sep=comma] {data/reeds/scalar_flux.csv};
		\addlegendentry{Reference}

	    \addplot [orange, thick, dashed, mark = square*] table [x = x, y = wls, col sep=comma] {data/reeds/scalar_flux.csv};
	    \addlegendentry{NDA WLS}

	    \addplot [teal, dashdotted, thick, mark = *] table [x = x, y = saaf, col sep=comma] {data/reeds/scalar_flux.csv};
	    \addlegendentry{NDA \saaft}

		\addplot [black, thin, below] coordinates {(2, 0) (2, 3)};
		\addplot [black, thin, below] coordinates {(3, 0) (3, 3)};
	    \addplot [black, thin, below] coordinates {(5, 0) (5, 3)};
	    \addplot [black, thin, below] coordinates {(6, 0) (6, 3)};

		\pgfplotsset{
			after end axis/.code={
				\node[black,below] at (axis cs:1,1.9){\small{\(\sigt[1] = 50\)}};
				\node[black,below] at (axis cs:1,1.7){\small{\(q_1 = 50\)}};

				\node[black,below] at (axis cs:2.5,1.9){\small{\(\sigt[2] = 5\)}};

				\node[black,below] at (axis cs:4,1.9){\small{\(\sigt[3] = 0\)}};

				\node[black,below] at (axis cs:5.5,0.7){\small{\(\sigt[4] = 1\)}};
		        \node[black,below] at (axis cs:5.5,0.5){\small{\(\sigs[4] = 0.9999\)}};
		        \node[black,below] at (axis cs:5.5,0.3){\small{\(q_4 = 1\)}};

		        \node[black,below] at (axis cs:7,2.7){\small{\(\sigt[5] = 1\)}};
		        \node[black,below] at (axis cs:7,2.5){\small{\(\sigs[5] = 0.9999\)}};
			}
		}

	\end{axis}
\end{tikzpicture}
        \end{minipage}
        \caption{Solution for the modified Reed's problem with NDA \saaft and NDA WLS. Comparison to a highly refined WLS reference solution. (Cross sections in \Xsunit and source strengths in \sourceunit).}
        \label{fig:reeds_results}
    \end{figure}
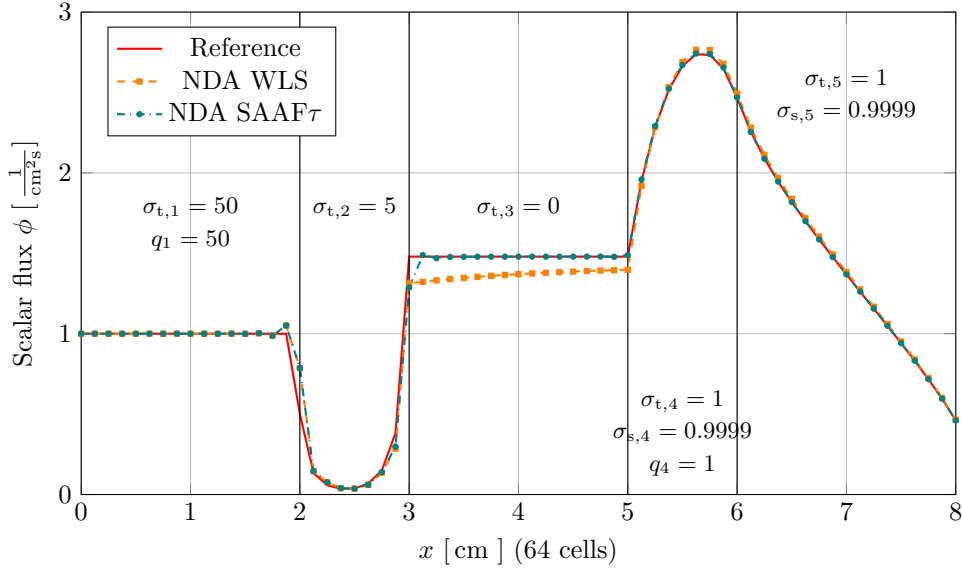

    \begin{figure}[th]
        \begin{minipage}{0.8\textwidth}
            \setlength{\figheight}{8cm}
            \begin{tikzpicture}
    \begin{axis}[
            xmin = 0, xmax = 8,
            ymin = 0, ymax = 60,
            y tick label style={/pgf/number format/.cd, scaled y ticks = false, fixed},
            x tick label style={/pgf/number format/.cd, scaled x ticks = false, fixed},,
            xtick = {0, 1, ..., 8},
            height=\figheight,
            width=1.0\textwidth,
            grid=major,
       		legend pos = north east,
            legend columns = 1,
            xlabel = {\(x\) [\cm] (64 cells)},
            ylabel = {Relative scalar flux error [\%]},
            mark options={solid, scale = 0.4}
        ]

        \addplot [orange, solid, thick, mark = square*] table [x = x, y expr = {abs(\thisrow{wls} - \thisrow{reference}) / \thisrow{reference} * 100}, col sep=comma] {data/reeds/scalar_flux.csv};
        \addlegendentry{NDA WLS}

        \addplot [teal, dashed, thick, mark = *] table [x = x, y expr = {abs(\thisrow{saaf} - \thisrow{reference}) / \thisrow{reference} * 100}, col sep=comma] {data/reeds/scalar_flux.csv};
        \addlegendentry{NDA \saaft}

        \addplot [black, mark=none] coordinates {(2, 0) (2, 60)};
        \addplot [black, mark=none] coordinates {(3, 0) (3, 60)};
        \addplot [black, mark=none] coordinates {(5, 0) (5, 60)};
        \addplot [black, mark=none] coordinates {(6, 0) (6, 60)};

    \end{axis}
\end{tikzpicture}
        \end{minipage}
        \caption{Relative error for the modified Reed's problem with NDA \saaft and NDA WLS to the WLS reference solution.}
        \label{fig:reeds_error_rel}
    \end{figure}
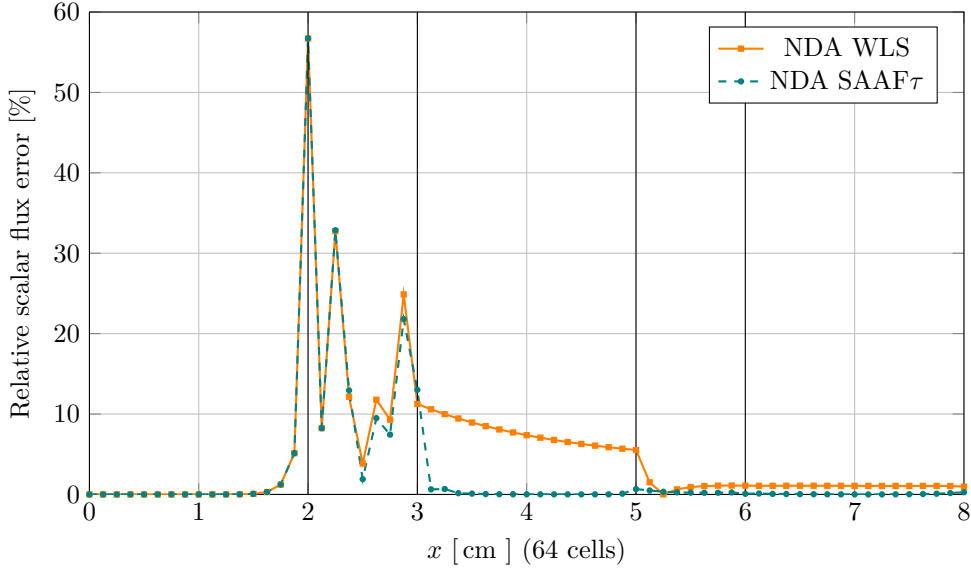

    To test the void NDA modifications, we used a slightly modified version of Reed's problem, a well known test problem containing a void region and a highly diffusive region. The Rattlesnake calculations used both NDA schemes: the NDA WLS and the NDA \saaft~\cite{wang_diffusion_2014}. The results are shown in \cref{fig:reeds_results} and the relative error in the scalar flux can be seen in \cref{fig:reeds_error_rel}. Both schemes have large errors in the absorber region. These are mainly due to by the rapid variation of the scalar flux in that region. In the void region the NDA WLS solution shows a non-constant flux and an incorrect magnitude. This affects the adjacent scattering region. The NDA \saaft solution showed small oscillations at the void's left boundary and a decrease in the scalar flux only in the leftmost cell in the void. These inaccuracies in both NDA WLS and NDA \saaft disappear with increasing mesh refinement. Both schemes needed 16 nonlinear Picard iterations to reduce the error between two consecutive low order solutions below the relative error tolerance of \tento{-10}. \par

    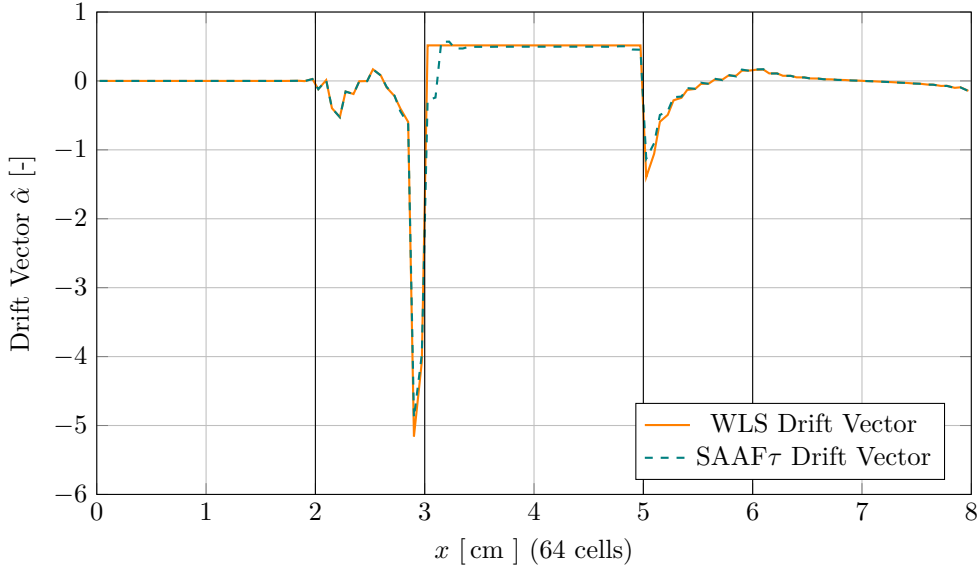
\begin{figure}[th]
        \begin{minipage}{0.8\textwidth}
            \setlength{\figheight}{8cm}
            \newsavebox{\detail}

\savebox{\detail} {
    \begin{tikzpicture}
    \begin{axis}[
            ticklabel style = {font=\scriptsize},
            axis line style={draw=none},
            axis background/.style={fill=gray!10},
            xmin = 2.95, xmax = 3.5,
            ymin = -0.8, ymax = 0.65,
            xtick = {3.0,3.1,...,3.5},
            height=0.65\figheight,
            width=0.6\textwidth,
            grid=major,
        ]

        \addplot [red, thick] table [x = x, y = ls_1000, col sep=comma] {data/reeds/drift_vectors.csv};
        \addplot [blue, thick, dashed] table [x = x, y = saaf_orig, col sep=comma] {data/reeds/drift_vectors.csv};

    \end{axis}
    \end{tikzpicture}
}

\begin{tikzpicture}
    \begin{axis}[
        xmin = 0, xmax = 8,
            ymin = -6.0, ymax = 1.0,
            y tick label style={/pgf/number format/.cd, scaled y ticks = false, fixed},
            x tick label style={/pgf/number format/.cd, scaled x ticks = false, fixed},,
            xtick = {0, 1, ..., 8},
            ytick = {-6, -5, ..., 1},
            height=\figheight,
            width=1.0\textwidth,
            grid=major,
            legend pos = south east,
            legend columns = 1,
            cycle list name = color list,
            xlabel = {\(x\) [\cm] (64 cells)},
            ylabel = {Drift Vector \drift[] [-]}
        ]

        \addplot [orange, thick] table [x = x, y = ls_1000, col sep=comma] {data/reeds/drift_vectors.csv};
        \addlegendentry{WLS Drift Vector}

        \addplot [teal, thick, dashed] table [x = x, y = saaf_orig, col sep=comma] {data/reeds/drift_vectors.csv};
        \addlegendentry{\saaft Drift Vector}

        \addplot [black] coordinates {(2, -10) (2, 1)};
        \addplot [black] coordinates {(3, -10) (3, 1)};
        \addplot [black] coordinates {(5, -10) (5, 1)};
        \addplot [black] coordinates {(6, -10) (6, 1)};

        \pgfplotsset{
            after end axis/.code={
            }
        }
    \end{axis}
\end{tikzpicture}
        \end{minipage}
        \caption{Drift vectors from the WLS and SAAF transport calculation for the modified Reed's problem.}
        \label{fig:reeds_drift}
    \end{figure}

    The drift vectors from both WLS and \saaft agree well as can be seen in \cref{fig:reeds_drift}, except for the left cells in the void region. The \saaft drift vector has oscillation on the left side of the void region. The WLS drift vector is constant throughout the void region. \par

\subsection{Two region problem with void}

\newcommand{\xL}{x_\mathrm{L}}
\newcommand{\xI}{x_\mathrm{I}}
\newcommand{\xR}{x_\mathrm{R}}

    To further investigate the non-constant flux of the NDA WLS solution in the void region as shown in \cref{fig:reeds_results} we simplified the problem to an one dimensional two region problem. The left half of the problem contains a void (\(\sigt[1] = 0\Xsunit\)), while the right side contains a strong absorber (\(\sigt[2] = 10\Xsunit\)). On the left boundary is an incident isotropic flux \(\phi^\mathrm{inc} = 1.0\sfluxunit\). The problem is 2\cm wide with \(\xL\) the left boundary, \(\xR\) the right boundary and \(\xI = 1\cm\) the interface between the void and the absorber. We study this pure absorber problem even though no NDA iterations or acceleration are required to obtain the solution. However, the low-order equation also ensures conservation for the WLS scheme, so it is reasonable to use it in the case of zero or small scattering ratios. Additionally, we avoided feedback from the low-order equation into the transport equation because of no scattering and were able to isolate the effects of the low-order equation. \par
    
    This problem uses a mesh with a large grid size. We picked the coarse mesh for two reasons, first to make the problem obvious. The same problem occurs in finer meshes or with smaller cross sections, but the error is much smaller. Using a coarse mesh makes the error big enough to be seen easily. The second reason is, that it is of interest how a method behaves for coarse cells, because in large geometries not all interfaces can be appropriately resolved.

    It is easy to obtain an analytic solution to this simple problem~\cite{hammer_nonlinear_2017}. The analytical solution in the void with an isotropic incoming flux on the left boundary is given by
    \begin{subequations} \label{eq:void_analytic_solution}
        \begin{equation}
            \phi_1\left(x\right) = \frac{\phi^\mathrm{inc}}{2} \label{eq:void_analytic_1}
        \end{equation}
        where the subscript 1 stands for the void left half of the problem and for the absorption region
        \begin{equation}
            \phi_2\left(x\right) = \frac{\phi^\mathrm{inc}}{2} \mathrm{E}_2\left(\sigt\left(x - \xI\right)\right) \label{eq:void_analytic_2}
        \end{equation}
    \end{subequations}
    with subscript 2. \(\mathrm{E}_n\) is the exponential-integral function
    \begin{equation} \label{eq:exponential_integral_fct}
        \mathrm{E}_n\left(x\right) \equiv \int_{1}^{\infty} \frac{\e{-xt}}{t^n} \dx[t].
    \end{equation}
     \par

    With the analytic solution it is possible to calculate the drift vector \drift analytically
    \begin{subequations} \label{eq:appendix_void_drift_analytic}
        \begin{align} \label{eq:appendix_void_drift_analytic_1}
            \drift_1\left(x\right) &= - \frac{1}{2} \\
            \label{eq:appendix_void_drift_analytic_2}
            \drift_2\left(x\right) &= -\frac{1}{\mathrm{E}_2\left(\sigt\left(x - \xI\right)\right)} \left(
            \left[\frac{1}{2}\mathrm{E}_3 \left(\sigt\left(x - \xI\right)\right)\right] - \DC_2\left(x\right)\left[\mathrm{E}_1\left(\sigt\left(x - \xI\right)\right)\right]\right).
        \end{align}
    \end{subequations}
    The analytic non-local diffusion coefficient has the form
    \begin{subequations} \label{eq:appendix_void_nldc}
        \begin{align} \label{eq:appendix_void_nldc_region_1}
        \DC_1\left(x\right) 
        &= \frac{\xI - \xL}{2} + \frac{1}{2\sigt}\left(\frac{1}{3} - \mathrm{E}_4\left(\sigt\left(\xR - \xI\right)\right)\right) \\
        \label{eq:appendix_void_nldc_region_2}
        \DC_2\left(x\right) 
        &= \frac{1}{3\sigt} 
        - \frac{1}{2\sigt}\big(\mathrm{E}_4\left(\sigt\left(\xR - x\right)\right) + \mathrm{E}_4\left(\sigt\left(x - \xI\right)\right)\big) 
        + \frac{\xI - \xL}{2}\mathrm{E}_3\left(\sigt\left(x - \xI\right)\right)
        \end{align}
    \end{subequations}
    where the void part is only dependent on the width of the void plus the boundary inflows. For the absorber region the coefficient is the classical diffusion coefficient with a correction for boundary effects. A \sn analytic solution, employing Gauss-quadrature to integrate over the angle was used as a reference for the transport solutions and the corresponding NDA solutions. \par

    Clearly in the void region the drift vector \(\drift_1 = -0.5\) and the non-local diffusion coefficient \DC[1] are constant. Therefore, the drift-diffusion equation \cref{eq:drift_diffusion} can be simplified to
    \begin{equation} \label{eq:void_drift_diffusion}
        -\DC_1 \ddxx \phi_1 - \drift_1 \ddx \phi_1 = 0
    \end{equation}
    The analytical solution to \cref{eq:void_drift_diffusion} is
    \begin{equation} \label{eq:void_drift_diffusion_solution}
        \phi_1 \left(x\right) = A_1 + B_1\e{-\frac{\drift_1}{\DC_1}x}
    \end{equation}
    with \(A_1\) and \(B_1\) constants to be determined by the boundary and interface conditions. As we can see, the constant solution is part of the solution space of \cref{eq:void_drift_diffusion_solution} but not the exclusive one. For a nonzero constant \(B_1\) the solution can also be exponential. \par

    
    The problem does not feature scattering, hence no iteration process is required to obtain the solution. This allowed us to compare the transport solution and the solutions to the drift-diffusion equation using different drift vectors. These drift vectors were obtained from the analytical solution (\cref{eq:appendix_void_drift_analytic}) and from the WLS and \saaft transport solve.

    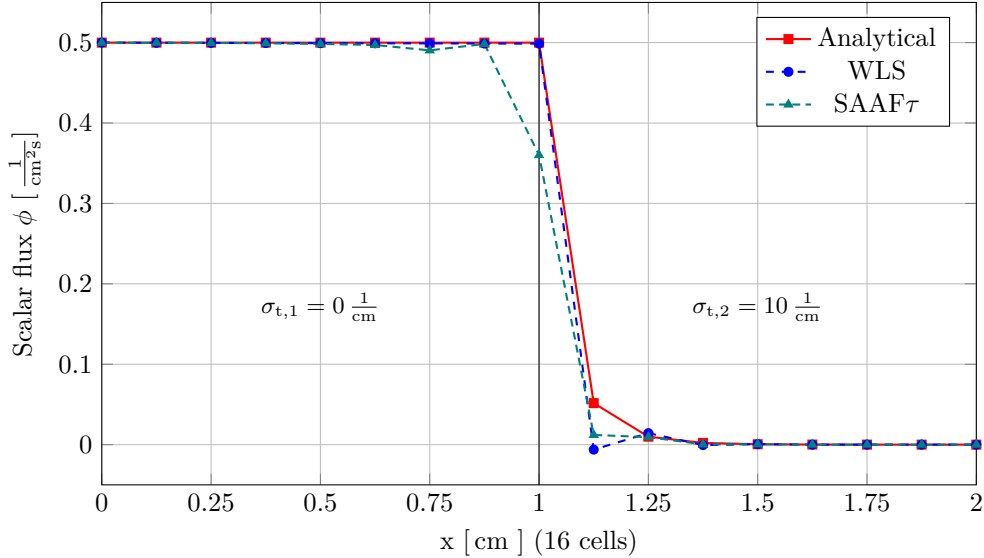
\begin{figure}[th]
        \begin{minipage}{0.8\textwidth}
            \setlength{\figheight}{8cm}
            \begin{tikzpicture}
  \begin{axis}[
        xmin = 0.0, xmax = 2,
        ymin = -0.05, ymax = 0.55,
        xtick = {0.0,0.25,...,2.0},
        ytick = {0.0,0.1,...,1.0},
        height=\figheight,
        width=\textwidth,
        grid=major,
        legend pos = north east,
        legend columns = 1,
        xlabel = {x [\cm] (16 cells)},
        ylabel = {Scalar flux \(\phi\) [\sfluxunit]},
        mark options={solid, scale = 0.75}
    ]

    \addplot [red, thick, mark = square*] table [x = x, y = sn_analytic, col sep=comma, each nth point={3}] {data/void/analytic_nonlocal_n_8.csv};
    \addlegendentry{Analytical}

    \addplot [blue, dashed, thick, mark = *] table [x = x, y = ls, col sep=comma, each nth point={3}] {data/void/analytic_nonlocal_n_8.csv};
    \addlegendentry{WLS}

    \addplot [teal, densely dashed, thick, mark = triangle*] table [x = x, y = saaf, col sep=comma, each nth point={3}] {data/void/analytic_nonlocal_n_8.csv};
    \addlegendentry{SAAF\(\tau\)}

    \addplot [black, thin, below] coordinates {(1, -0.1) (1, 1.5)};

	\pgfplotsset{
        after end axis/.code={
            \node[black,below] at (axis cs:0.5,0.2){\small{\(\sigt[1] = 0\Xsunit\)}};
            \node[black,below] at (axis cs:1.5,0.2){\small{\(\sigt[2] = 10\Xsunit\)}};
        }
    }

  \end{axis}
\end{tikzpicture}
        \end{minipage}
        \caption{WLS and \saaft transport solutions for the two region problem with a void and an incident isotropic flux on the left side compared to an analytic reference solution.}
        \label{fig:void_transport}
    \end{figure}

    \Cref{fig:void_transport} shows the solution to the problem using the WLS and \saaft transport solvers in comparison to the analytic \sn[8] solution. The WLS used a weight function limit of \(\weight[\mathrm{max}] = 1000\cm\) and the \saaft used \(\zeta = 0.5\). These parameters were also used for the remaining results in this section. The result of the WLS scheme showed a constant flux in the void region. The \saaft scheme started oscillating towards the right side of the void region and dropped significantly in the last cell before the material interface. Both schemes had a dip after the interface in the material half and continuing oscillations into the material region, which is a typical behavior on material interfaces of second order equations. \par

    Now that the angular fluxes of the transport solutions were known, we were able to calculate all correction terms for the NDA. As mentioned before, with the correction terms the low order drift-diffusion solution can be obtained without any further transport solve. Hence, we can compare the results for the different drift vectors without any feedback from the drift-diffusion solution, which we would have, if we were required to iterate. \par

    \begin{figure}[th]
        \begin{minipage}{0.8\textwidth}
            \setlength{\figheight}{8cm}
            \begin{tikzpicture}
  \begin{axis}[
        xmin = 0.0, xmax = 2,
        ymin = -0.25, ymax = 0.9,
        xtick = {0.0,0.25,...,2.0},
        ytick = {0.0,0.25,0.5,0.75,1.0},
        height=\figheight,
        width=\textwidth,
        grid=major,
        legend pos = north east,
        legend columns = 1,
        xlabel = {x [\cm] (16 cells)},
        ylabel = {Scalar flux \(\phi\) [\sfluxunit]},
        mark options={solid, scale = 0.75}
    ]

    \addplot [red, thick] table [x = x, y = analytic, col sep=comma, each nth point={3}] {data/void/analytic_nonlocal_n_8.csv};
    \addlegendentry{Analytical}

    \addplot [orange, dashed, thick, mark = square*] table [x = x, y = CFEM sn_analytic, col sep=comma, each nth point={3}] {data/void/analytic_nonlocal_n_8.csv};
    \addlegendentry{NDA \sn Analytic}

    \addplot [blue, densely dashed, thick, mark = *] table [x = x, y = CFEM wls, col sep=comma, each nth point={3}] {data/void/analytic_nonlocal_n_8.csv};
    \addlegendentry{NDA WLS}

    \addplot [teal, solid, thick, mark = triangle*] table [x = x, y = CFEM saaf_tau, col sep=comma, each nth point={3}] {data/void/analytic_nonlocal_n_8.csv};
    \addlegendentry{NDA SAAF\(\tau\)}

    \addplot [black, below] coordinates {(1, -1) (1, 1.5)};

    \pgfplotsset{
        after end axis/.code={
            \node[black,below] at (axis cs:0.5,0.25){\small{\(\sigt[1] = 0\Xsunit\)}};
            \node[black,below] at (axis cs:1.5,0.25){\small{\(\sigt[2] = 10\Xsunit\)}};
        }
    }
  \end{axis}
\end{tikzpicture}
        \end{minipage}
        \caption{NDA solutions to the two region problem with a void and an incident isotropic flux on the left side using different drift vectors.}
        \label{fig:void_nda}
    \end{figure}
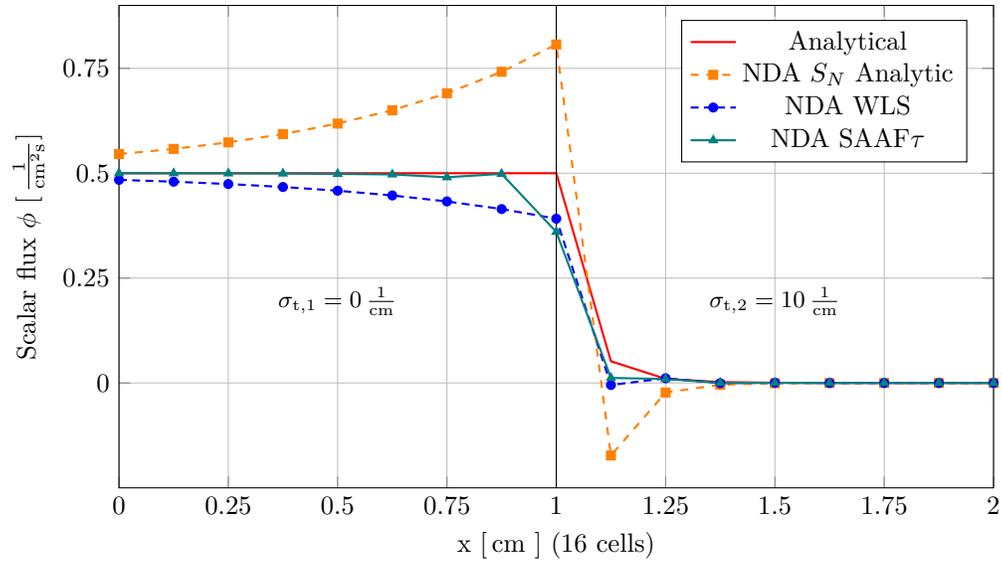

    \begin{figure}[th]
        \begin{minipage}{0.8\textwidth}
            \setlength{\figheight}{8cm}
            \begin{tikzpicture}
  \begin{semilogyaxis}[
        xmin = 0.0, xmax = 2,
        ymin = 1e-4, ymax = 1e3,
        xtick = {0.0,0.25,...,2.0},
        ytick = {1e-4,1e-3,1e-2,1e-1,1e0, 1e1, 1e2, 1e3},
        height=\figheight,
        width=\textwidth,
        grid=major,
    	legend pos = south east,
        legend columns = 2,
        xlabel = {x [\cm] (16 cells)},
        ylabel = {Relative scalar flux error [\%]},
        mark options={solid, scale = 0.75}
    ]

    \addplot [red, solid, thick, mark = square*] table [x = x, y expr = {abs(\thisrow{ls} - \thisrow{sn_analytic}) / \thisrow{sn_analytic} * 100}, col sep=comma, each nth point={3}] {data/void/analytic_nonlocal_n_8.csv};
    \addlegendentry{WLS}

    \addplot [purple, solid, thick, mark = *] table [x = x, y expr = {abs(\thisrow{saaf} - \thisrow{sn_analytic}) / \thisrow{sn_analytic} * 100}, col sep=comma, each nth point={3}] {data/void/analytic_nonlocal_n_8.csv};
    \addlegendentry{SAAF\(\tau\)}

    \addplot [orange, dashed, thick, mark = triangle*] table [x = x, y expr = {abs(\thisrow{CFEM sn_analytic} - \thisrow{sn_analytic}) / \thisrow{sn_analytic} * 100}, col sep=comma, each nth point={3}] {data/void/analytic_nonlocal_n_8.csv};
    \addlegendentry{NDA \sn Analytic}

    \addplot [blue, dotted, very thick, mark = diamond*] table [x = x, y expr = {abs(\thisrow{CFEM wls} - \thisrow{sn_analytic}) / \thisrow{sn_analytic} * 100}, col sep=comma, each nth point={3}] {data/void/analytic_nonlocal_n_8.csv};
    \addlegendentry{NDA WLS}

    \addplot [teal, dashed, very thick, mark = halfcircle*] table [x = x, y expr = {abs(\thisrow{CFEM saaf_tau} - \thisrow{sn_analytic}) / \thisrow{sn_analytic} * 100}, col sep=comma, each nth point={3}] {data/void/analytic_nonlocal_n_8.csv};
    \addlegendentry{NDA SAAF\(\tau\)}

    \addplot [black, below] coordinates {(1, 1e-4) (1, 1000)};

  \end{semilogyaxis}
\end{tikzpicture}
        \end{minipage}
        \caption{Relative error in the scalar flux of the transport and NDA solutions for the two region problem with a void, the NDA solutions use different drift vector vectors.}
        \label{fig:void_error_rel}
    \end{figure}
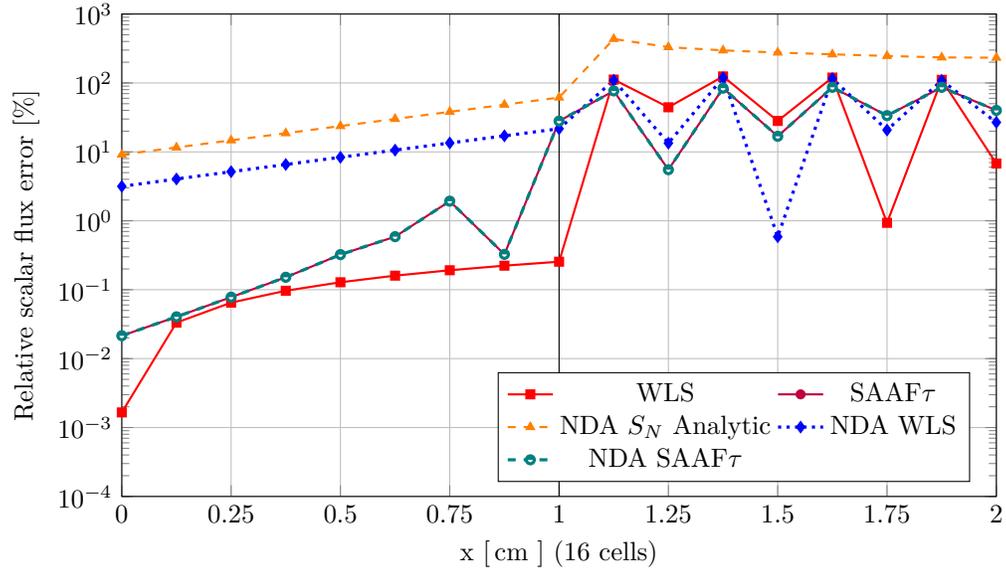

    If we use the drift vectors from the transport solves discussed above, we obtain the NDA solutions shown in \cref{fig:void_nda}. All calculation used an analytic expression for the non-local diffusion coefficient (\cref{eq:appendix_void_nldc}). The WLS drift vector produced an exponentially decreasing flux in the void regions. As we have shown earlier, this is part of the solution space of the drift-diffusion equation. The drift vector obtained from the analytical solution gave the worst result. The scalar flux in the void region was exponentially increasing towards the material interface. The results show that a more accurate drift vector does not necessarily increase the accuracy of the NDA solution. As shown in \cref{eq:void_drift_diffusion_solution} the interface and boundary conditions of the void region determine the shape of the scalar flux within the void. The \saaft was constant in the left part of the void region. In the last cell before the interface it decreased strongly. The \saaft drift vector is not constant in voids, hence \cref{eq:void_drift_diffusion,eq:void_drift_diffusion_solution} are not valid. The oscillations of the transport solution forced the NDA solution to be consistent. \par

    The relative error in the scalar flux is shown in \cref{fig:void_error_rel}. The largest error showed the NDA with the analytic drift vector. In the void region the WLS transport solution had the least error. The \saaft transport and the NDA using the \saaft drift vector had the same error as expected because of the consistency. All numerical schemes showed approximately the same error in the material region with strong oscillations. \par

    \begin{figure}[p]
        \begin{minipage}{0.8\textwidth}
            \setlength{\figheight}{8cm}
            \begin{tikzpicture}
    \begin{loglogaxis}[
       		xmin = 1, xmax = 5e4,
       		ymin = 1e-8, ymax = 1,
            ytick = {1e-8,1e-6,1e-4,1e-2,1},
            y tick label style={/pgf/number format/.cd, scaled y ticks = false, fixed},
            x tick label style={
                xticklabel={
                    \pgfkeys{/pgf/fpu=true}
                    \pgfmathparse{exp(\tick)}%
                    \pgfmathprintnumber[fixed relative, precision=3]{\pgfmathresult}
                    \pgfkeys{/pgf/fpu=false}
                }
            },
            height=\figheight,
            width=\textwidth,
            grid=major,
            legend pos = south west,
            legend columns = 1,
            cycle list name = exotic,
            xlabel = {Number of spatial cells [-]},
            ylabel = {Relative \(\mathrm{L}_2\) error [-]},
            mark options={solid}
        ]

        \addplot [red, thick, mark = square*] table [x = n, y = ls, col sep=comma] {data/void/analytic_nonlocal_error_norm.csv};
        \addlegendentry{WLS}

        \addplot [blue, dashed, thick, mark = *] table [x = n, y = saaf, col sep=comma] {data/void/analytic_nonlocal_error_norm.csv};
        \addlegendentry{SAAF}

        \addplot [teal, dashdotted, thick, mark = triangle] table [x = n, y = CFEM sn_analytic, col sep=comma] {data/void/analytic_nonlocal_error_norm.csv};
        \addlegendentry{NDA \sn Analytic}

        \addplot [orange, dotted, very thick, mark = square] table [x = n, y = CFEM ls, col sep=comma] {data/void/analytic_nonlocal_error_norm.csv};
        \addlegendentry{NDA WLS}

        \addplot [magenta, dashdotdotted, thick, mark = o] table [x = n, y = CFEM saaf, col sep=comma] {data/void/analytic_nonlocal_error_norm.csv};
        \addlegendentry{NDA SAAF\(\tau\)}

    \end{loglogaxis}
\end{tikzpicture}
        \end{minipage}
        \caption{Convergence of the error with mesh refinement for the two region problem for transport and NDA solutions.}
        \label{fig:void_convergence}
    \end{figure}
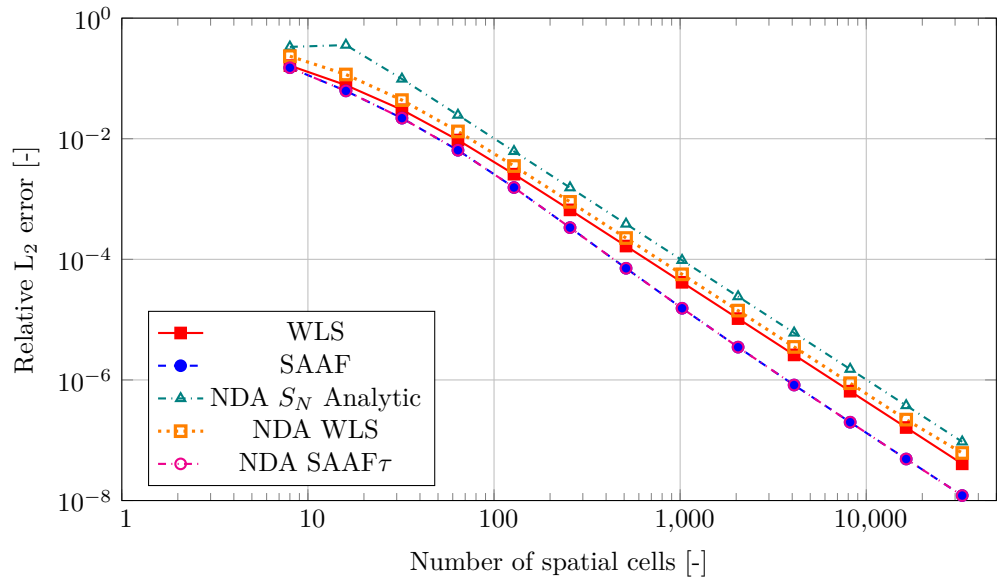

    The described error in the void is a coarse mesh problem. Increasing refinement of the mesh reduced the error as shown in \cref{fig:void_convergence}. All schemes converged spatially with second order. \par

    \begin{figure}[p]
        \begin{minipage}{0.8\textwidth}
            \setlength{\figheight}{8cm}
            \begin{tikzpicture}
    \begin{loglogaxis}[
       		xmin = 1, xmax = 5e4,
       		ymin = 1e-8, ymax = 1,
            ytick = {1e-8,1e-6,1e-4,1e-2,1},
            y tick label style={/pgf/number format/.cd, scaled y ticks = false, fixed},
            x tick label style={
                xticklabel={
                    \pgfkeys{/pgf/fpu=true}
                    \pgfmathparse{exp(\tick)}%
                    \pgfmathprintnumber[fixed relative, precision=3]{\pgfmathresult}
                    \pgfkeys{/pgf/fpu=false}
                }
            },
            height=\figheight,
            width=\textwidth,
            grid=major,
            legend pos = south west,
            legend columns = 1,
            cycle list name = exotic,
            xlabel = {Number of spatial cells in material region [-]},
            ylabel = {Relative \(\mathrm{L}_2\) error [-]},
            mark options={solid}
        ]

        \addplot [red, thick, mark = square*] table [x = n, y = ls, col sep=comma] {data/void/analytic_nonlocal_error_norm_constant_void.csv};
        \addlegendentry{WLS}

        \addplot [blue, dashed, thick, mark = *] table [x = n, y = saaf, col sep=comma] {data/void/analytic_nonlocal_error_norm_constant_void.csv};
        \addlegendentry{SAAF}

        \addplot [teal, dashdotted, thick, mark = triangle] table [x = n, y = CFEM sn_analytic, col sep=comma] {data/void/analytic_nonlocal_error_norm_constant_void.csv};
        \addlegendentry{NDA \sn Analytic}

        \addplot [orange, dotted, very thick, mark = square] table [x = n, y = CFEM wls, col sep=comma] {data/void/analytic_nonlocal_error_norm_constant_void.csv};
        \addlegendentry{NDA WLS}

        \addplot [magenta, dashdotdotted, thick, mark = o] table [x = n, y = CFEM saaf_tau, col sep=comma] {data/void/analytic_nonlocal_error_norm_constant_void.csv};
        \addlegendentry{NDA SAAF\(\tau\)}

    \end{loglogaxis}
\end{tikzpicture}
        \end{minipage}
        \caption{Convergence of the error for the two region problem for transport and NDA solutions with constant 8 cells in the void region and mesh refinement in the material region.}
        \label{fig:void_convergence_const}
    \end{figure}
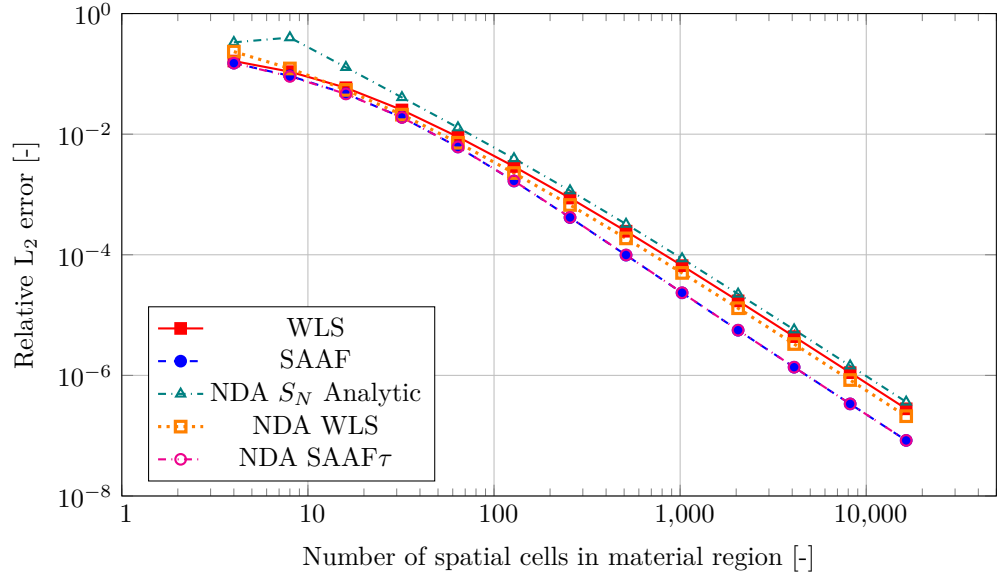

    We also studied the spatial convergence if we kept the number of cells in the void constant (8 cells). Since void regions normally do not hold many details, refinement might be a waste of computational resources. The results in \cref{fig:void_convergence_const} showed, that the spatial convergence is second order to the number of cells in the material region. This indicates again, that the error in the void region is caused by the error on the material interface propagated into the void. Improving the error in the material regions hence also improves the error in the void region. \par

    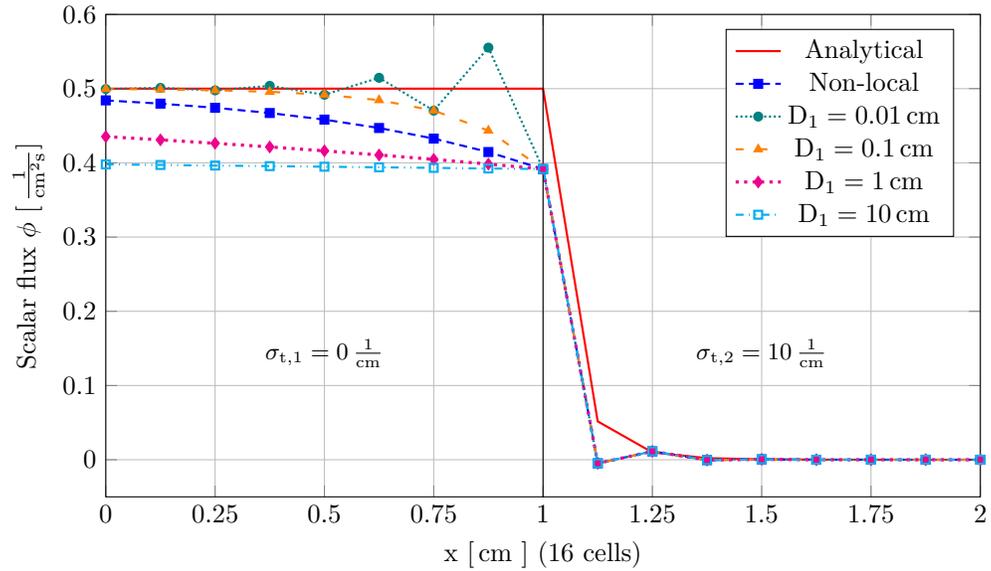
\begin{figure}[tp]
        \begin{minipage}{0.8\textwidth}
            \setlength{\figheight}{8cm}
            \begin{tikzpicture}
  \begin{axis}[
        xmin = 0.0, xmax = 2,
        ymin = -0.05, ymax = 0.6,
        xtick = {0.0,0.25,...,2.0},
        ytick = {0.0,0.1,...,1.0},
        height=\figheight,
        width=\textwidth,
        grid=major,
        legend pos = north east,
        legend columns = 1,
        xlabel = {x [\cm] (16 cells)},
        ylabel = {Scalar flux \(\phi\) [\sfluxunit]},
        mark options={solid, scale = 0.75}
    ]

    \addplot [red, thick] table [x = x, y = sn_analytic, col sep=comma, each nth point={3}] {data/void/max_d_100.csv};
    \addlegendentry{Analytical}

    \addplot [blue, densely dashed, thick, mark = square*] table [x = x, y = CFEM wls, col sep=comma, each nth point={3}] {data/void/analytic_nonlocal_n_8.csv};
    \addlegendentry{Non-local}

    \addplot [teal, densely dotted, thick, mark = *] table [x = x, y = CFEM wls, col sep=comma, each nth point={3}] {data/void/max_d_0.01.csv};
    \addlegendentry{\(\DC[1] = 0.01\cm\)}

    \addplot [orange, loosely dashed, thick, mark = triangle*] table [x = x, y = CFEM wls, col sep=comma, each nth point={3}] {data/void/max_d_0.1.csv};
    \addlegendentry{\(\DC[1] = 0.1\cm\)}

    \addplot [magenta, dotted, very thick, mark = diamond*] table [x = x, y = CFEM wls, col sep=comma, each nth point={3}] {data/void/max_d_1.0.csv};
    \addlegendentry{\(\DC[1] = 1\cm\)}

    \addplot [cyan,, dashdotdotted, thick, mark = square] table [x = x, y = CFEM wls, col sep=comma, each nth point={3}] {data/void/max_d_10.csv};
    \addlegendentry{\(\DC[1] = 10\cm\)}

    \addplot [black, below] coordinates {(1, -1) (1, 1.5)};

    \pgfplotsset{
        after end axis/.code={
            \node[black,below] at (axis cs:0.5,0.175){\small{\(\sigt[1] = 0\Xsunit\)}};
            \node[black,below] at (axis cs:1.5,0.175){\small{\(\sigt[2] = 10\Xsunit\)}};
        }
    }

  \end{axis}
\end{tikzpicture}
        \end{minipage}
        \caption{Results for the NDA WLS scheme using different diffusion coefficients in the void region.}
        \label{fig:void_nda_diffusion}
    \end{figure}

    Finally we were interested in the performance of the non-local diffusion coefficient compared to other diffusion coefficients. We used the local diffusion coefficient with several constant values in the void region and compared the results to the non-local diffusion coefficient in \cref{fig:void_nda_diffusion}. For the NDA WLS the choice of the diffusion coefficient has a large impact on the scalar flux in the void region. The scheme is independently differenced for small \sigt, hence small differences between the transport and NDA solution arise. If the diffusion coefficient is too large, these differences will be magnified and lead to an incorrect result in the void region (\(\DC \ge 10\cm\)). For small diffusion coefficients the result started to oscillate in the void region. The non-local diffusion coefficient \(\DC[nl] \approx 0.25\cm\) is, in this case, of the right magnitude. However, it is not the optimal choice to minimize the error. The non-local diffusion coefficient gives reasonable results, but a smaller diffusion coefficient reduced the error in the void. However, this small diffusion coefficient might have a negative effect on the iterative convergence for cases with scattering. \par

\FloatBarrier

\subsection{C5G7 reactor physics benchmark}

    The C5G7 MOX benchmark problem is a test without spatial homogenization for modern deterministic transport codes. We focused on the two dimensional version of the benchmark, for purpose of investigating the performance of all the schemes. The twenty sets of results that were initially submitted to the benchmark committee can be found in a special issue of Progress in Nuclear Energy~\cite{smith_benchmark_2004}. More recent calculations of the benchmark with a spatial and angular convergence study were presented by McGraw~\cite{mcgraw_accuracy_2015} and Wang~\cite{wang_convergence_2015}. \par
    
    We modified the C5G7 benchmark to test the WLS with or without NDA for voids. We ran two cases with water and graphite moderator. All guide tubes and the central fission chamber of each fuel assembly was converted into a void. To quantify the effect of single changes, calculations were also run with moderator but no voids further denoted by the graphite and water case respectively The case with voids and graphite moderator will be referenced as C\_void case, the case with voids and water moderator as H2O\_void. The result of all cases were compared against PDT~\cite{hawkins_efficient_2012} calculations provided by McGraw. There are small differences between the water cases and the original benchmark, the central fission chambers were replaced by water. The PDT reference was obtained with the same mesh size and angular quadrature. \par

\begin{figure}
        \includegraphics*[scale=1]{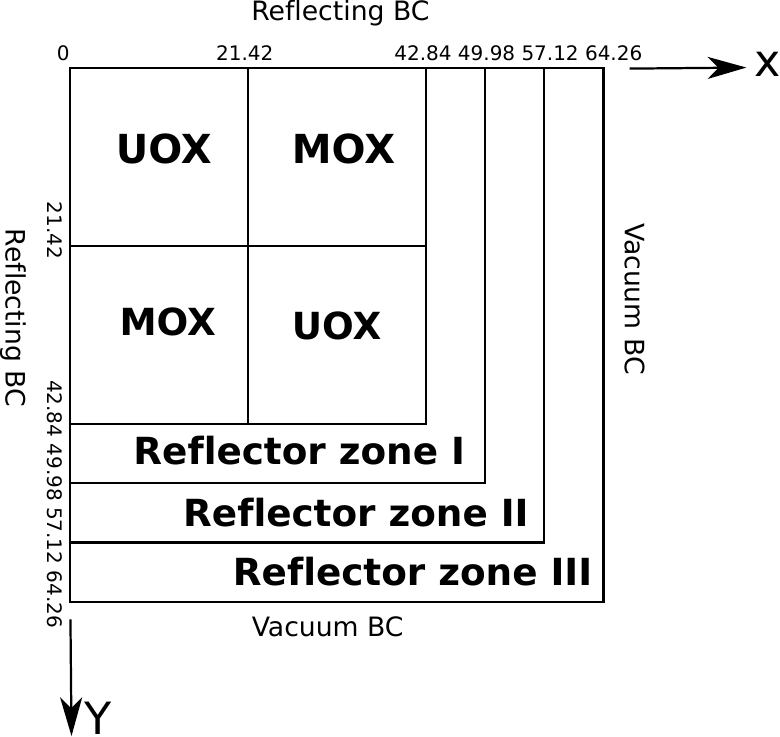}
        \caption{Zone layout of the C5G7 benchmark geometry~\cite{wang_convergence_2015}.}
        \label{fig:c5g7_geomtry}
    \end{figure}

    The mesh was generated using the 2D mesh generator Triangle~\cite{shewchuk_triangle:_1996} with a geometry file which is created by a Rattlesnake mesh generator. The quality of the mesh ensures that no triangle has an interior angle less than 20 degrees. In order to limit the number of elements in the mesh, the surrounding reflector region is divided into three separate regions as shown in \cref{fig:c5g7_geomtry} employing a coarser mesh far away from the fuel region, while the same maximum triangle area is applied to all fuel assemblies.\par

    For the calculations we use the mesh with 8 equal sides approximating the circumference of the fuel pins as described by Wang \cite{wang_convergence_2015}. The mesh conserve the volume of each fuel pin and hence the mass of fissile material. \par

    The implementation of the algorithm in \rattlesnake is given in a different paper~\cite{hammer_weighted_2018}. This paper also presented results for the WLS and \saaft schemes with corresponding NDA for the original benchmark. \par

    \rattlesnake provides routines to calculate the non-local diffusion coefficient~\cite{schunert_using_2016}. Two options are available, the on-the-fly calculation and the prepared calculation. The on-the-fly calculation provides the non-local diffusion tensor on every quadrature point in the domain. The auxiliary transport system is solved separately from the main solve and any transport scheme provided by \rattlesnake can be used. The system for the diffusion tensor is automatically set up and solved prior to the main solve. This method provides the highest accuracy, however it requires the auxiliary system to be solved before every calculation. This can be expensive for real world problems. \par

    The second method generates the non-local diffusion tensors in a completely separated calculation and writes the resulting diffusion tensors into the cross section file. This method eliminates the need to recalculate the diffusion tensor for every run. The limitation of this methods is, that the information are only available per homogenized region, so with far less detail than the on-the-fly calculations. Nevertheless we chose to use this method. The reduction in computational time and the reduced amount of memory necessary to perform the calculations allowed us to actual run the problem without an excessive amount of computational resources. Another reason is that the accuracy of the non-local diffusion coefficient, if sufficiently high, has only a limited influence on the result. \par

    These calculations were performed with Gauss-Chebychev quadrature with 4 polar and 32 azimuthal angles per octant. For the transport calculations a relative tolerance of \tento{-8} on L2 norm of residual was used. The NDA calculations used as convergence criteria the difference between successive scalar flux iterates with a threshold of \tento{-8} with a high order relative tolerance of \tento{-4}. These tolerances were used for all following calculations. \par

    The first thing to establish is the effect of the non-local diffusion coefficient on the results. The non-local coefficient is necessary for the void calculations, nevertheless it can also be used for the cases without voids. The local and the non-local diffusion coefficient can hence be compared. \Cref{tab:c5g7_graphite_ls} shows the comparison for the errors in the eigenvalue and the average and maximal pin power errors. The use of the non-local diffusion coefficient only affected the result if the NDA scheme was independently differenced (NDA LS). The reason for this is, that high-order and low-order solution for these schemes were not exactly the same, consequently the diffusion terms in \cref{eq:drift_vector,eq:drift_diffusion} do not cancel. Nevertheless, the associated error was approximately 1\,pcm or less than 0.1\,\% for the pin powers. The dependently differenced schemes showed no difference between the calculations using the local coefficient and the ones using the non-local diffusion coefficient. \par
    
    \begin{table}
        \caption{Comparison of the local and the non-local diffusion coefficient results for the eigenvalue error and the average and maximal pin power error for the C5G7 graphite case.}
        \label{tab:c5g7_graphite_ls}
        \begin{tabular}{lcc@{\hskip 35pt}cc@{\hskip 35pt}cc}
        	\toprule
        	Scheme     &           \multicolumn{2}{c@{\hskip 35pt}}{\keff error [pcm]}           &            \multicolumn{2}{c@{\hskip 35pt}}{Avg. error [\%]}            &            \multicolumn{2}{c}{Max. error [\%]}            \\
        	           & \multicolumn{1}{c}{local} & \multicolumn{1}{c@{\hskip 35pt}}{non-local} & \multicolumn{1}{c}{local} & \multicolumn{1}{c@{\hskip 35pt}}{non-local} & \multicolumn{1}{c}{local} & \multicolumn{1}{c}{non-local} \\ \midrule
        	NDA LS     &          70.024           &                   71.104                    &           0.063           &                    0.064                    &           0.815           &             0.820             \\
        	NDA WLS    &          68.653           &                   68.653                    &           0.068           &                    0.068                    &           0.869           &             0.869             \\
        	NDA SAAF   &          68.653           &                   68.653                    &           0.068           &                    0.068                    &           0.869           &             0.869             \\
        	NDA \saaft &          17.316           &                   17.316                    &           0.071           &                    0.071                    &           0.520           &             0.520             \\ \bottomrule
        \end{tabular}
    \end{table}

    \begin{table}[tp]
        \caption{Eigenvalues for all C5G7 calculations: Original (original benchmark geometry) Graphite (graphite moderator), void (graphite moderator and voids), Water (water moderator), Void2 (water moderator and voids).}
        \label{tab:append_c5g7_k_eff}
        \setlength{\tabcolsep}{15pt}
            \begin{tabular}{lrrrrr}
                \toprule
                Scheme     & Original & Graphite &  C\_void &   Water & H2O\_void \\ \midrule
                PDT        &  1.18646 &  0.65639 & 0.65405 & 1.19862 &  1.17829 \\
                LS         &  1.34527 &  0.65657 & 0.65466 & 1.35451 &  1.34532 \\
                WLS        &  1.18558 &  0.65570 & 0.65320 & 1.19845 &  1.17682 \\
                SAAF       &  1.18558 &  0.65570 &       - & 1.19845 &        - \\
                \saaft     &  1.18698 &  0.65622 & 0.65391 & 1.19963 &  1.17890 \\ \midrule
                NDA LS     &  1.18593 &  0.65568 & 0.65369 & 1.19900 &  1.17835 \\
                NDA WLS    &  1.18558 &  0.65570 & 0.65349 & 1.19845 &  1.17740 \\
                NDA SAAF   &  1.18558 &  0.65570 &       - & 1.19845 &        - \\
                NDA \saaft &  1.18698 &  0.65622 & 0.65391 & 1.19963 &  1.17890 \\ \bottomrule
            \end{tabular}
        \setlength{\tabcolsep}{6pt}
    \end{table}

    \begin{table}[p]
        \caption{Errors in \keff and the average and maximal error in the pin powers for the graphite (graphite moderator and no voids) and the C\_void case (graphite moderator and voids) of the modified C5G7 benchmark.}
        \label{tab:c5g7_void_1}
        \begin{tabular}{lrr@{\hskip 35pt}rr@{\hskip 35pt}rr}
        	\toprule
        	Scheme     &           \multicolumn{2}{c@{\hskip 35pt}}{\keff error [pcm]}           &            \multicolumn{2}{c@{\hskip 35pt}}{Avg. error [\%]}            &            \multicolumn{2}{c}{Max. error [\%]}            \\
        	           & \multicolumn{1}{c}{graphite} & \multicolumn{1}{r@{\hskip 35pt}}{C\_void} & \multicolumn{1}{c}{graphite} & \multicolumn{1}{r@{\hskip 35pt}}{C\_void} & \multicolumn{1}{c}{graphite} & \multicolumn{1}{r}{C\_void} \\ \midrule
        	LS         &                       18.348 &                                   61.775 &                        0.372 &                                    0.444 &                        1.978 &                      2.118 \\
        	WLS        &                       68.653 &                                   84.537 &                        0.068 &                                    0.097 &                        0.869 &                      1.043 \\
        	\saaft     &                       17.316 &                                   13.268 &                        0.071 &                                    0.075 &                        0.520 &                      0.476 \\ \midrule
        	NDA LS     &                       71.104 &                                   35.854 &                        0.064 &                                    0.120 &                        0.820 &                      0.461 \\
        	NDA WLS    &                       68.653 &                                   55.401 &                        0.068 &                                    0.090 &                        0.869 &                      0.695 \\
        	NDA \saaft &                       17.316 &                                   13.268 &                        0.071 &                                    0.075 &                        0.520 &                      0.476 \\ \bottomrule
        \end{tabular}
    \end{table}

    \begin{table}[p]
        \caption{Errors in \keff and the average and maximal error in the pin powers for the water (water moderator and no voids) and the void2 case (water moderator and voids) of the modified C5G7 benchmark.}
        \label{tab:c5g7_void_water_1}
        \begin{tabular}{lrr@{\hskip 35pt}rr@{\hskip 35pt}rr}
            \toprule
            Scheme     &          \multicolumn{2}{c@{\hskip 35pt}}{\keff error [pcm]}           &           \multicolumn{2}{c@{\hskip 35pt}}{Avg. error [\%]}            &           \multicolumn{2}{c}{Max. error [\%]}            \\
            & \multicolumn{1}{c}{water} & \multicolumn{1}{r@{\hskip 35pt}}{H2O\_void} & \multicolumn{1}{c}{water} & \multicolumn{1}{r@{\hskip 35pt}}{H2O\_void} & \multicolumn{1}{c}{water} & \multicolumn{1}{r}{H2O\_void} \\ \midrule
            LS         &                 15589.399 &                                  16702.573 &                     8.474 &                                      8.631 &                    31.698 &                       33.012 \\
            WLS        &                    16.420 &                                    147.175 &                     0.470 &                                      0.435 &                     2.594 &                        2.080 \\
            \saaft     &                   101.149 &                                     60.980 &                     0.307 &                                      0.306 &                     1.752 &                        1.718 \\ \midrule
            NDA LS     &                    38.233 &                                      5.631 &                     0.467 &                                      0.543 &                     2.765 &                        2.844 \\
            NDA WLS    &                    16.420 &                                     89.650 &                     0.470 &                                      0.551 &                     2.594 &                        2.773 \\
            NDA \saaft &                   101.149 &                                     60.980 &                     0.307 &                                      0.306 &                     1.752 &                        1.718 \\ \bottomrule
        \end{tabular}
    \end{table}

    After we established that the non-local diffusion coefficient has only a minimal effect on the independently differenced LS schemes, we proceeded to the case containing voids. The introduction of the voids instead of graphite in all guide tubes and the central rod of each fuel assembly had only a small influence on the eigenvalue of the problem. The reference solution showed a difference of 234\,pcm between the two cases. \Cref{tab:append_c5g7_k_eff} shows a comparison of all eigenvalues. The comparison of the errors to the PDT solution are shown in \cref{tab:c5g7_void_1,tab:c5g7_void_water_1}. \cref{tab:c5g7_void_1} shows the error in eigenvalue and the average and maximal pin power errors for the graphite cases. \par
    
    The error in the eigenvalue for the non-conservative LS transport schemes is surprisingly small compared to the large errors seen in the original benchmark~\cite{hammer_weighted_2018} and the water cases (\cref{tab:c5g7_void_water_1}). We suspect that it is caused by the error cancellation. This theory is supported by the large average and maximal pin power errors. These were significantly larger than the errors for the other schemes. The WLS transport scheme showed the largest error in the eigenvalue for both graphite cases, however for the pure water case it showed the lowest. The errors for WLS scheme, which is conservative for cases with sufficient large cross sections increased approximately by one quarter with the introduction of voids. The best transport scheme was \saaft. It also did not show a significant increase in errors when voids were introduced into the problem. \par

\begin{figure}[p]
        \begin{minipage}{\textwidth}
            \includegraphics[width=0.95\textwidth]{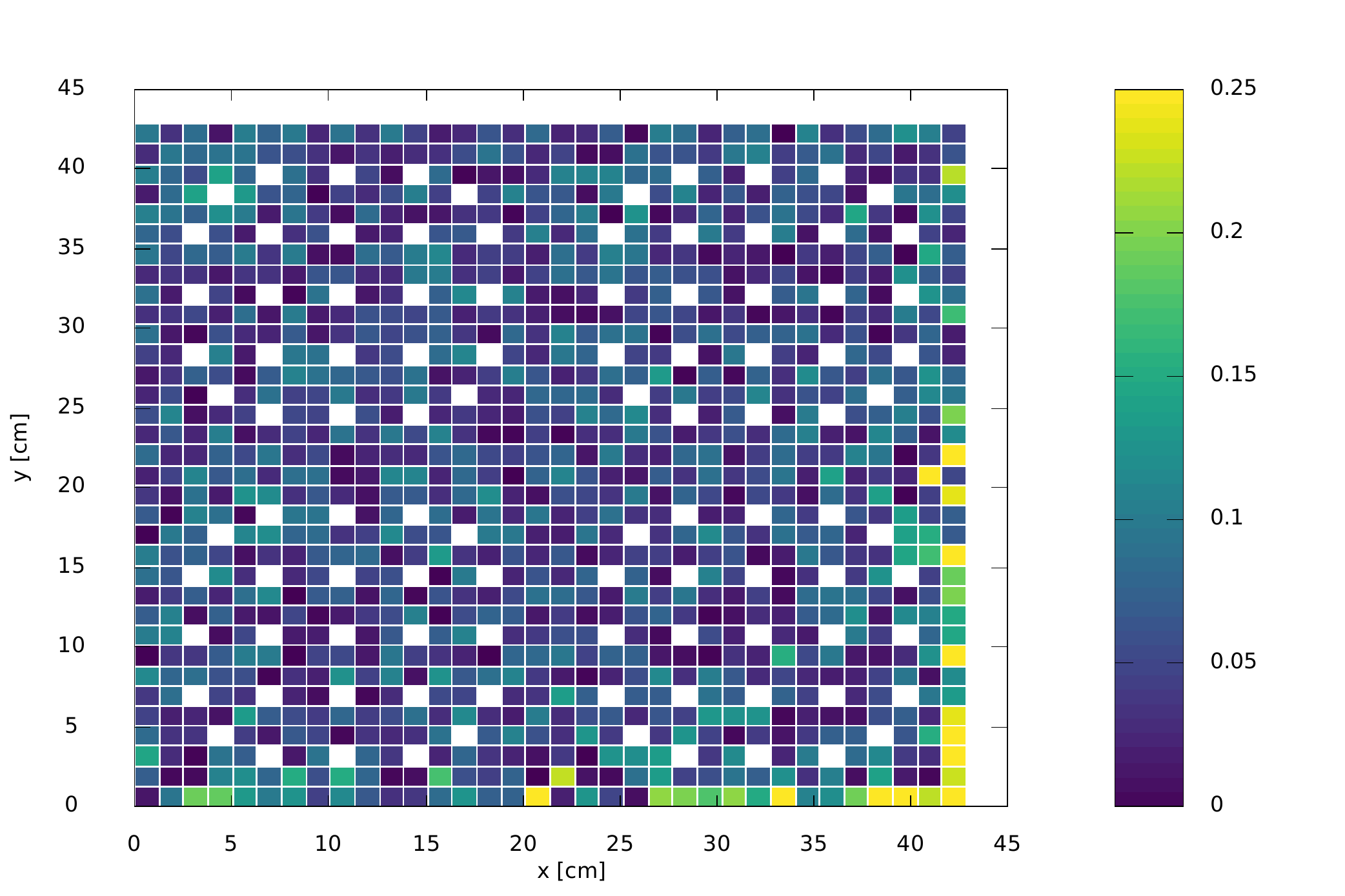}
            \subcaption{Graphite case}
        \end{minipage}
        \begin{minipage}{\textwidth}
            \includegraphics[width=0.95\textwidth]{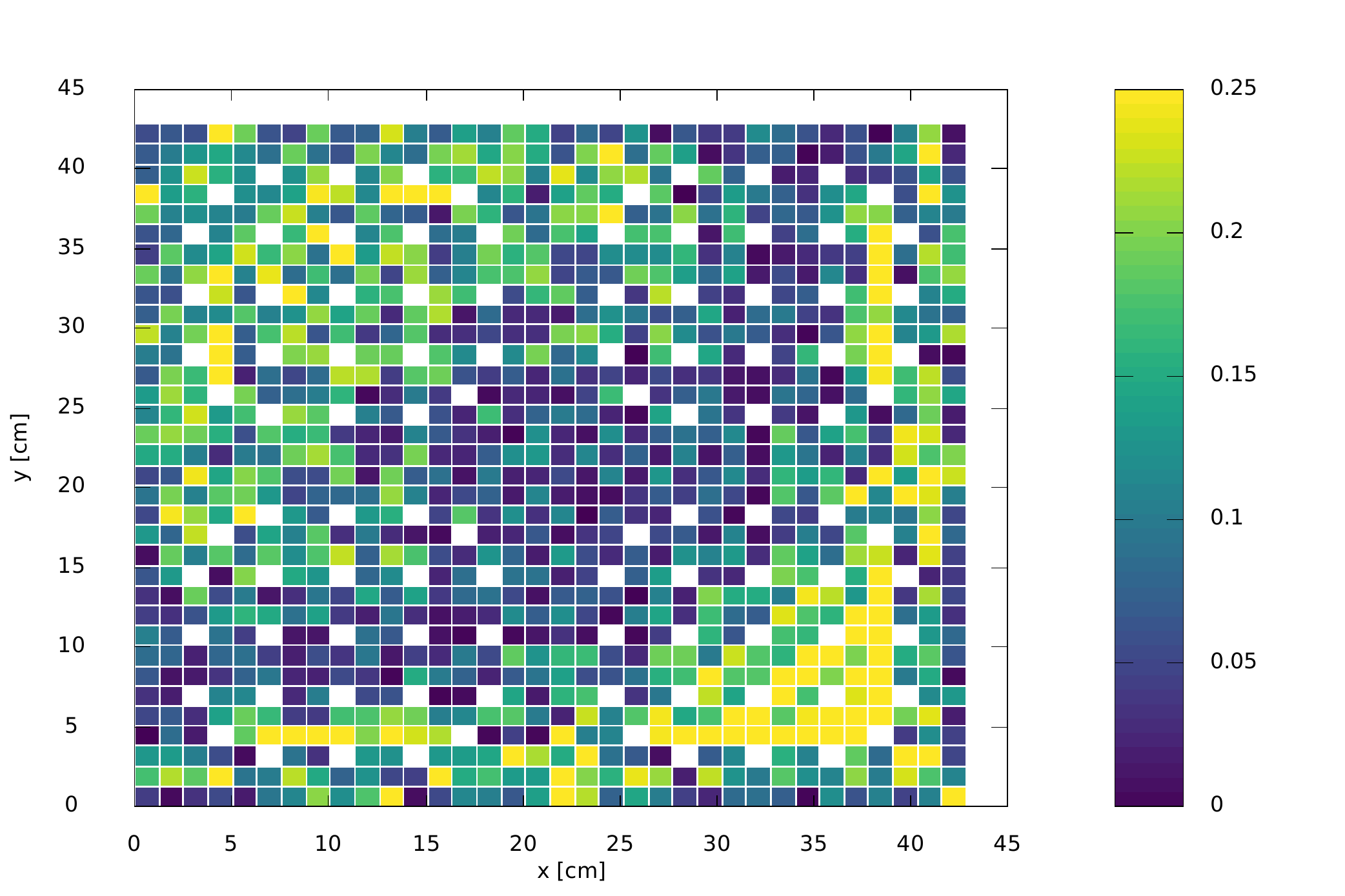}
            \subcaption{Void case}
        \end{minipage}
        \caption{Distribution of the pin power errors in percent for the NDA LS scheme, scale limited to 0.25\,\%.}
        \label{fig:append_c5g7_errors_ls_1}
    \end{figure}

\begin{figure}[p]
        \begin{minipage}{\textwidth}
            \includegraphics[width=0.95\textwidth]{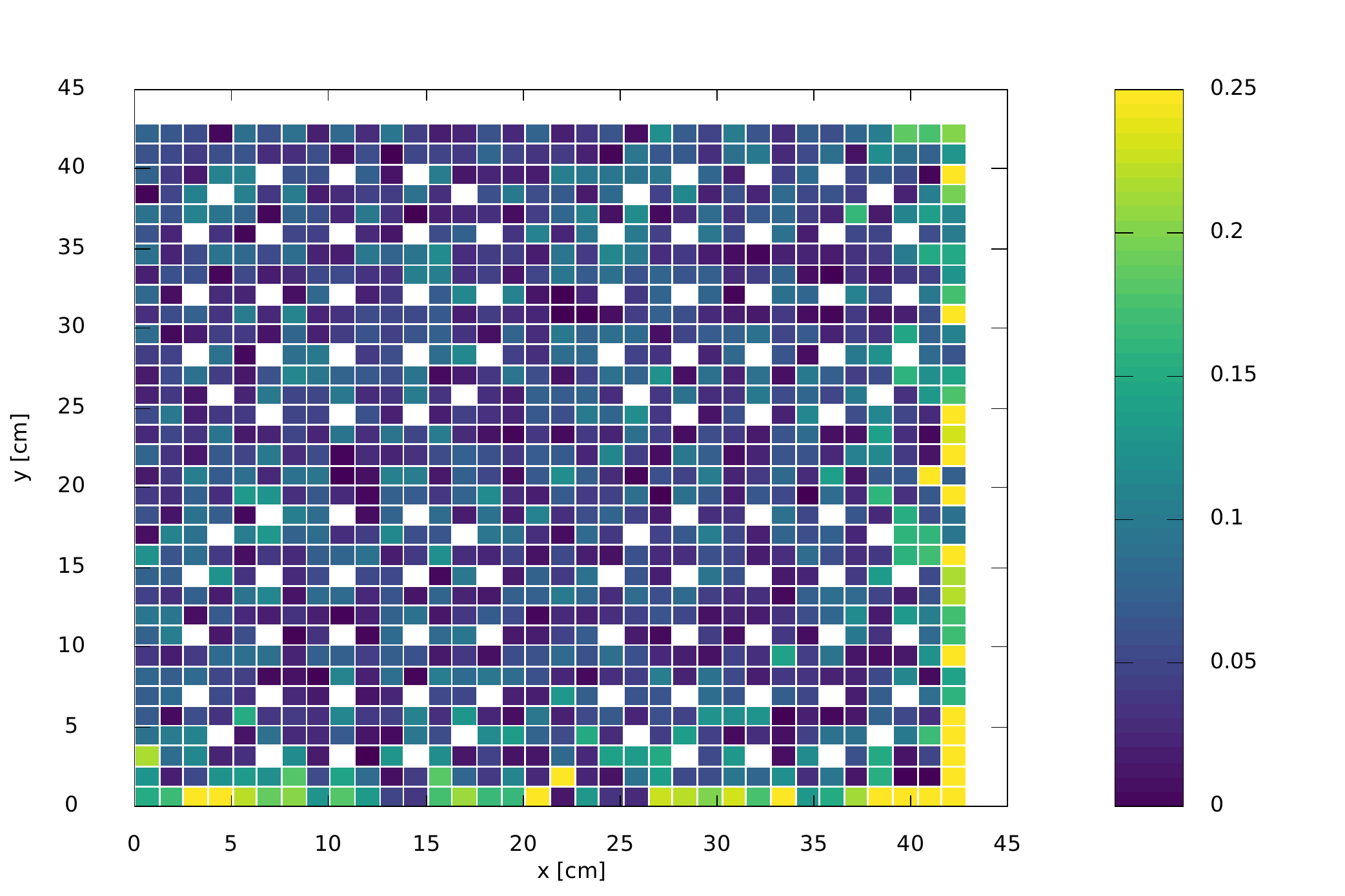}
            \subcaption{Graphite case}
        \end{minipage}
        \begin{minipage}{\textwidth}
            \includegraphics[width=0.95\textwidth]{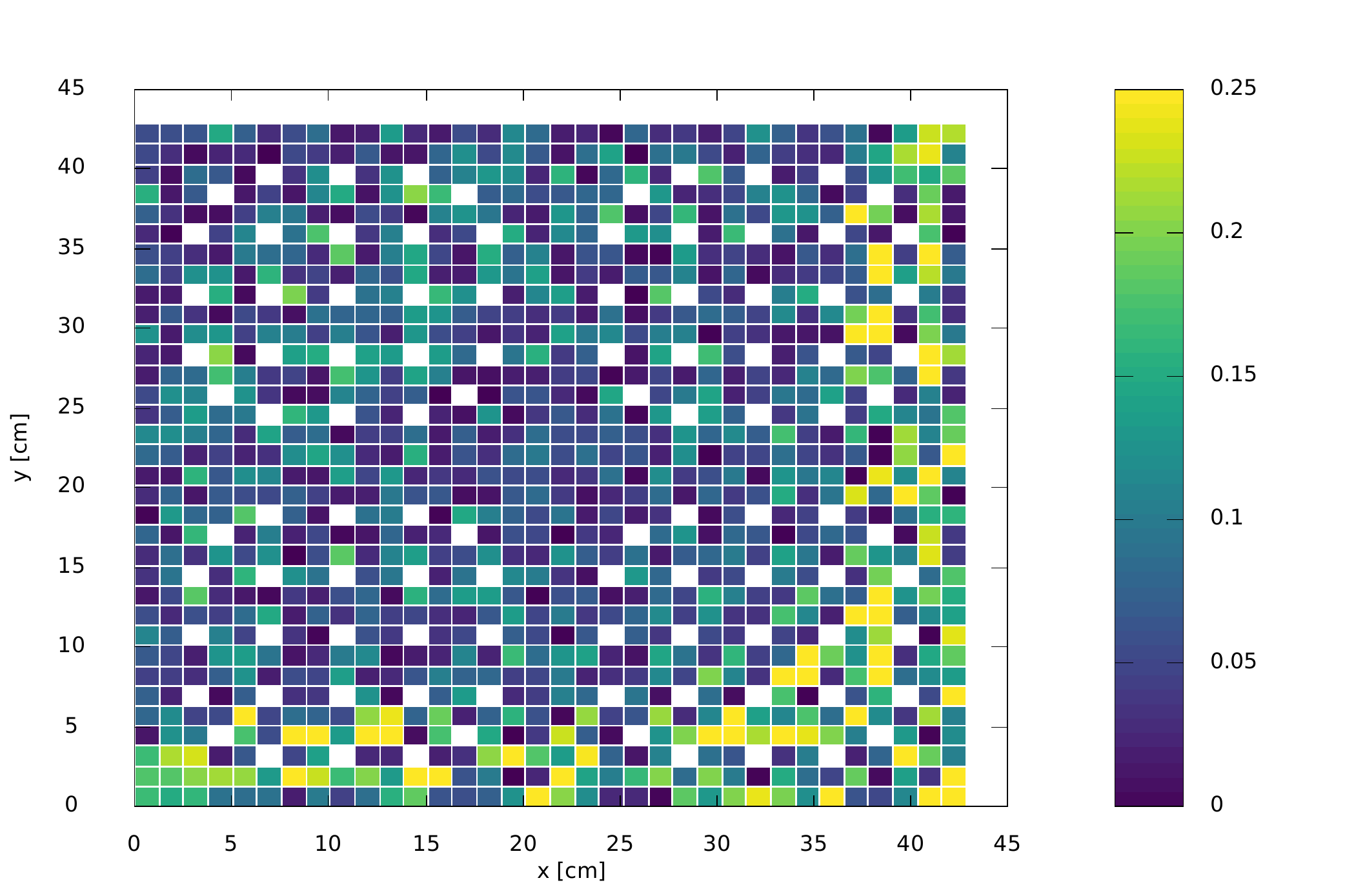}
            \subcaption{Void case}
        \end{minipage}
        \caption{Distribution of the pin power errors in percent for the NDA WLS scheme, scale limited to 0.25\,\%.}
        \label{fig:append_c5g7_errors_wls_1}
    \end{figure}

\begin{figure}[p]
        \begin{minipage}{\textwidth}
            \includegraphics[width=0.95\textwidth]{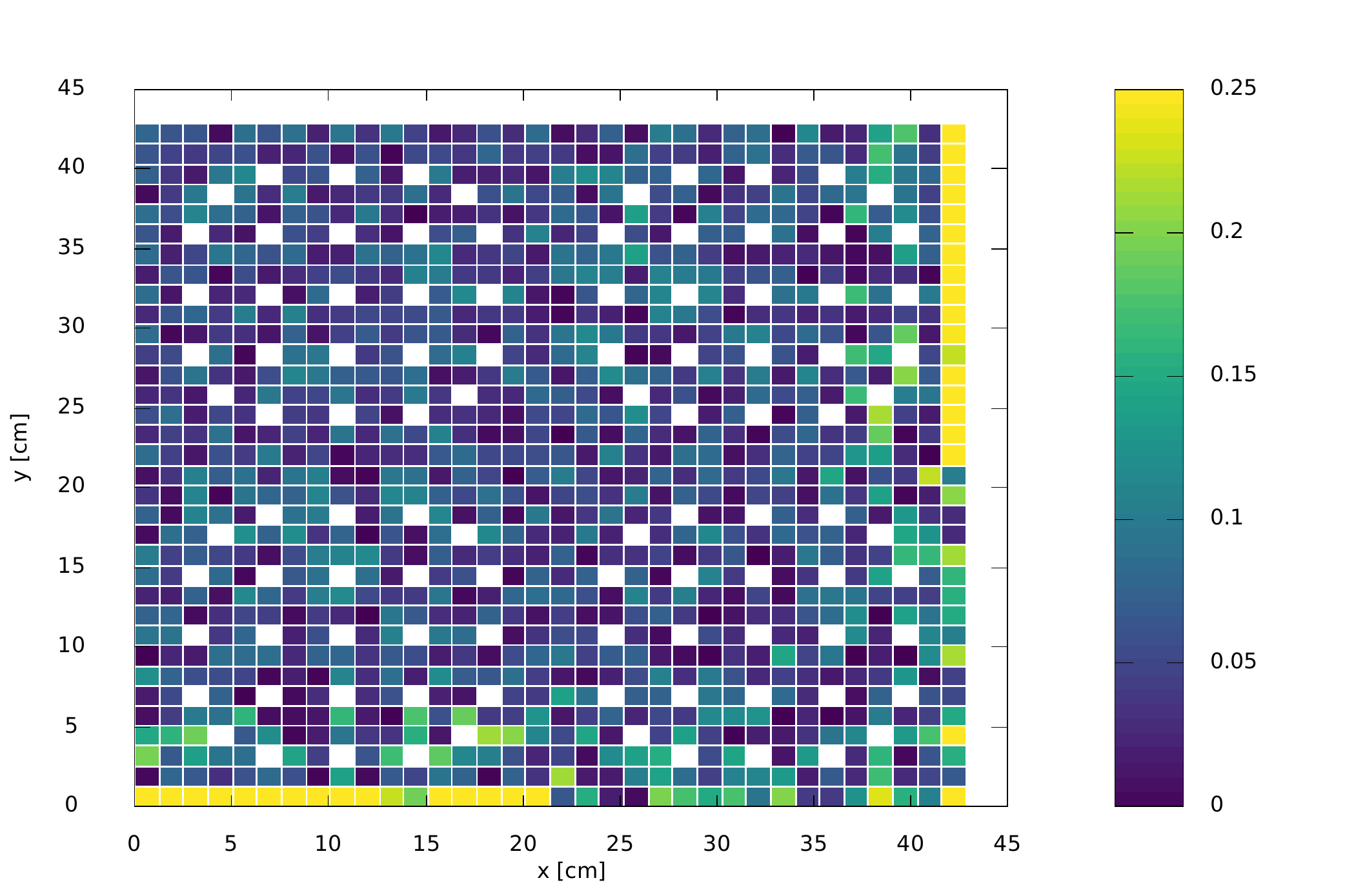}
            \subcaption{Graphite case}
        \end{minipage}
        \begin{minipage}{\textwidth}
            \includegraphics[width=0.95\textwidth]{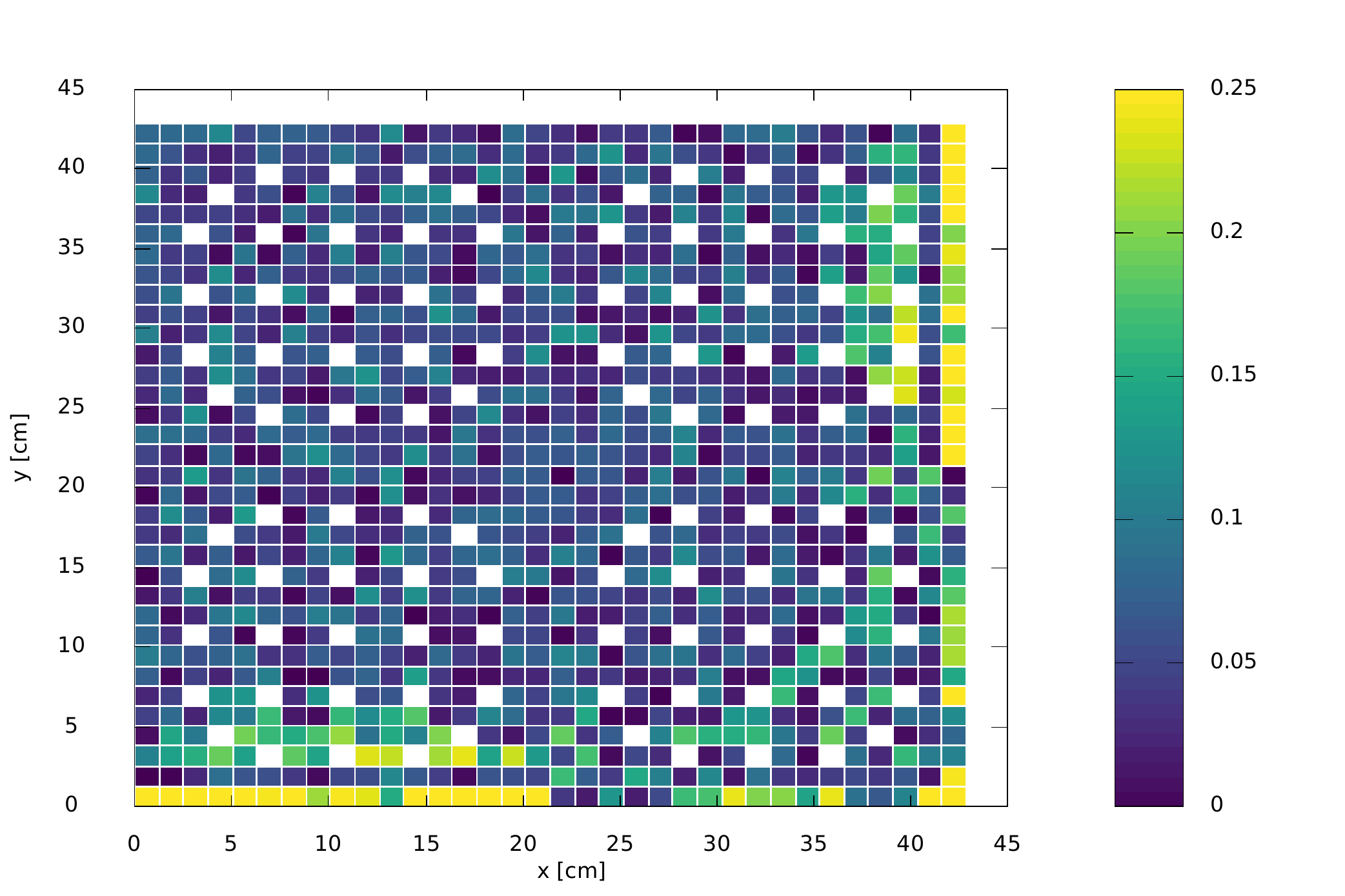}
            \subcaption{Void case}
        \end{minipage}
        \caption{Distribution of the pin power errors in percent for the NDA \saaft scheme, scale limited to 0.25\,\%.}
        \label{fig:append_c5g7_errors_saaft}
    \end{figure}

    The NDA schemes for LS and WLS are conservative even with geometry containing voids. This can clearly seen in the results for the error in \keff. The NDA LS schemes error in the eigenvalue was larger than the pure transport but the pin power errors are now comparable with the NDA WLS solution. Again we see error cancellation for void case in \keff for the LS schemes. The average pin power errors increase significantly, which can be seen in \cref{fig:append_c5g7_errors_ls_1}. While for the graphite case the error was limited to the pins close to the reflector region, in the void case large errors occurred in the central fuel element. Note that for the error plots the scale is limited to 0.25\,\% to show the distribution of the errors better; the error in single fuel pins can be larger than this, especially in the lower right corner pin. The plots have the same orientation as \cref{fig:c5g7_geomtry}. For the NDA WLS scheme the error in \keff decreased but the average error in the pin power increased. \Cref{fig:append_c5g7_errors_wls_1} show that for the graphite cases the error was located in the pins close to the graphite reflector. For the void case these errors stretch further towards the center of the core. Additionally, fuel elements close to a void tube showed larger errors than for the graphite case. These increases were smaller than for the LS cases. The \saaft NDA is consistently differenced with the low-order equation even in the void case. This scheme showed the lowest errors from all schemes. \Cref{fig:append_c5g7_errors_saaft} shows that in the void case the errors close to the reflector continued further inwards, however no increase in the center of the core can be seen. \par
    
    To demonstrate, that the non-conservative LS method has a large error, we resubstituted water as moderator instead of graphite. The results in \cref{tab:c5g7_void_water_1} show the large errors of the non-conservative schemes. Even the WLS scheme, which is nonconservative only in the voids and near-voids, showed a large error in the eigenvalue. The remaining results are similar to the results with graphite moderator. \par
    
     \par
    
    The condition of the system of the WLS scheme can be affected significantly by \(\weight[\mathrm{max}]\), which is indicated by the increasing CPU time or number of iterations in an iterative solver for the transport update. These results and the comparison on the accuracy are given in \cref{tab:c5g7_void_weight}.

    \begin{table}
        \caption{Comparison of the eigenvalue, pin power errors (average (AVG), mean relative error (MRE) and maximal error (MAX)) and relative runtime for increasing WLS weight function limit \(\weight[\mathrm{max}]\) of the NDA WLS scheme for the C\_void case (graphite moderator and voids).}
        \label{tab:c5g7_void_weight}
            \begin{tabular}{rrrrrrr}
            	\toprule
            	\weight[\mathrm{max}] &  \keff &   AVG &   MRE &   MAX & Relative runtime & Iterations \\
            	                 ~[-] &  [pcm] &  [\%] &  [\%] &  [\%] &     [-] &        [-] \\ \midrule
            	                    1 & 36.478 & 0.123 & 0.122 & 0.466 &     1.00 &         11 \\
            	                   10 & 46.382 & 0.096 & 0.096 & 0.591 &     1.09 &         11 \\
            	                  100 & 55.717 & 0.090 & 0.091 & 0.661 &     1.16 &         11 \\
            	                  500 & 56.685 & 0.092 & 0.093 & 0.660 &     1.45 &         11 \\
            	                 1000 & 56.809 & 0.092 & 0.094 & 0.660 &     2.93 &         11 \\ \bottomrule
        \end{tabular}
    \end{table}

    It is shown in \cref{tab:c5g7_void_weight} that the accuracy in \keff decreased with increasing \weight[\mathrm{max}], but the average pin power error improves. This is again a case of error cancellation for the eigenvalue as seen already above. The average pin power error did not improve much for \(\weight[\mathrm{max}] > 10\cm \). In graphical plots an improvement can be seen up to \(\weight[\mathrm{max}] = 100\cm\). For this \weight[\mathrm{max}] the runtime is still comparable to the NDA \saaft scheme. \par

\section{Conclusions}

    We derived a NDA algorithm based on our WLS transport equation which is well defined in geometries with voided regions. The drift vector was modified to use the direct formulation of the current for optically thin regions, while continuing to use the default Eddington formulation for optically thick cells. This combination gave the combined advantages of a well defined drift term in near voids and voids while maintaining the better convergence properties of the Eddington formulation for thick cells. The \(\tau\) formulation for NDA WLS proved itself not to be unconditionally stable for all cell thicknesses. \par    
    
    We employed a non-local definition of the diffusion coefficient, which is well defined in void regions by a transport solution of the surrounding geometry. This coefficient gives essentially a method to limit the diffusion coefficient based on the problem, while almost maintaining the local diffusion coefficient for optical thick regions. \par
    
    The better performance of the non-local diffusion coefficient in cases with high scattering ratios and the difficulties to define a problem dependent \DC[\mathrm{max}] convinced us to continue to use the non-local diffusion coefficient. It offers a method to limit the diffusion coefficient in voids automatically problem dependent. A wrong guess for \DC[\mathrm{max}] can strongly influence the convergence or make the problem even unstable. \par
    
    The numerical Fourier analysis confirmed that the void modifications are an unconditionally stable and efficient scheme. However they also showed that the calculation of the non-local diffusion coefficient can strongly influence the convergence rate, if the mesh for this calculations is unresolved and the non-local diffusion coefficient shows oscillations and negativities, that are caused by oscillations of the underlying WLS solve. Furthermore, the calculations of the non-local diffusion coefficient transforms any problem into a pure absorber problem, thus if a mesh is refined enough for a diffusive problem, it might not be sufficiently refined for the calculation of the non-local diffusion coefficient. Hence the non-local diffusion coefficient requires careful treatment to avoid instabilities in the NDA algorithm. \par
    
    The NDA results for the WLS showed non-constant behavior in void regions for slab geometries. We showed that this scalar flux solution is conservative and within the solution space of the drift-diffusion equation. Improved accuracy of the drift vector does not ameliorate the error in the void region, since it is caused by an unresolved boundary layer at the interface. \par
    
    We performed a study on pure absorber problems even though no NDA iterations are required because the NDA scheme enforces conservation. For these problems the diffusion coefficient is a free parameter. Any value can be chosen and a solution can be obtained. In problems with scattering, where acceleration is required, it is not a free parameter, because it can make the iteration scheme unstable. The scalar flux solution in the void for the NDA WLS is affected by the diffusion coefficient. The results with the non-local diffusion coefficient indicate that it is an acceptable choice with proven good iterative properties. \par
    
    The modified NDA scheme with non-local diffusion coefficient and the combined formulation of drift vectors was fully implemented in \rattlesnake. A modified C5G7 benchmark was used to test the new NDA scheme on a more complicated problem with voided regions. The comparison to PDT and NDA \saaft showed, that the results are reasonable accurate. While the \saaft NDA scheme was comparable in some cases or slightly better in others, it lacks the symmetric-positive definite properties of the NDA WLS scheme. Thus the NDA WLS scheme can use the conjugate-gradient method, which requires the storage of only three solutions vectors. Compared to GMRES, which can require an arbitrary number of solutions vectors or restarts with degraded convergence properties, this gives the NDA WLS scheme an enormous advantage regarding memory. Exploiting these advantages with \rattlesnake will be our future work. \par

\section{Acknowledgments}
    This material is based upon work supported by the Department of Energy, Battelle Energy Alliance, LLC, under Award Number DE-AC07-05ID14517.

\bibliographystyle{ans_js} 
\bibliography{literature}

\end{document}